\documentclass[a4paper,11pt]{article}
\pdfoutput=1 

\usepackage{jheppub} 
\makeatletter
\def\@fpheader{\relax}
\makeatother

\usepackage[T1]{fontenc} 

\usepackage{graphicx}
\usepackage{amsmath,amssymb,amsfonts}      
\usepackage{slashed}
\usepackage{hyperref}
\usepackage{graphicx,color}
\usepackage{colortbl}
\usepackage[usenames,dvipsnames,svgnames,table]{xcolor}
\definecolor{black}   {RGB}{0.0, 0.13, 0.28}
\definecolor{dukeblue}    {rgb}{0.0, 0.0, 0.61}
\usepackage{booktabs}

\usepackage{multirow}
\usepackage[capitalize]{cleveref}
\usepackage{xfrac}
\usepackage{cancel}
\usepackage{xspace}
\usepackage[compat=1.1.0]{tikz-feynman}
\tikzfeynmanset{warn luatex=false}

\newcommand{\smo}{\textsc{SModelS}\xspace}
\newcommand{\smodels}{\textsc{SModelS}\xspace}
\newcommand{\smothree}{\textsc{SModelS}~v3\xspace}

\newcommand{\Ztwo}{\ensuremath{{\mathcal Z}_2}}
\newcommand{\etmiss}{\ensuremath{{E_T^{\rm{miss}}}}}
\newcommand{\zp}{Z^{\prime}}

\newcommand{\zpm}{Z^{\prime \mu}}

\newcommand{\be}{\begin{equation}}
\newcommand{\ee}{\end{equation}}
\newcommand{\bi}{\begin{itemize}}
\newcommand{\ei}{\end{itemize}}

\newcommand{\up}{\text{U(1)}^{\prime}}
\newcommand{\code}[1]{\texttt{#1}}

\title{\boldmath SModelS v3: Going Beyond $\mathcal{Z}_2$ Topologies}

\author[a]{Mohammad~Mahdi~Altakach,}
\author[a]{Sabine~Kraml,}
\author[b]{Andre~Lessa,} 
\author[c,d]{Sahana~Narasimha,} 
\author[a]{Timoth\'ee~Pascal,} 
\author[b]{Camila~Ramos,} 
\author[b]{Yoxara~Villamizar,} 
\author[c,d]{Wolfgang~Waltenberger}  

\affiliation[a]{Laboratoire de Physique Subatomique et de Cosmologie (LPSC), Universit\'e
Grenoble-Alpes, \\ CNRS/IN2P3, 53 Avenue des Martyrs, F-38026 Grenoble, France}
\affiliation[b]{Centro de Ci\^encias Naturais e Humanas, Universidade Federal do ABC, Santo Andr\'e, 09210-580 SP, Brazil}
\affiliation[c]{Institut f\"ur Hochenergiephysik,  \"Osterreichische Akademie der Wissenschaften, Georg-Coch-Platz 2, A-1010 Wien, Austria}
\affiliation[d]{University of Vienna, Faculty of Physics, Boltzmanngasse 5, A-1090 Wien, Austria}

\emailAdd{altakach@lpsc.in2p3.fr}
\emailAdd{sabine.kraml@lpsc.in2p3.fr}
\emailAdd{andre.lessa@ufabc.edu.br}
\emailAdd{sahana.narasimha@oeaw.ac.at}
\emailAdd{pascal@lpsc.in2p3.fr}
\emailAdd{ramos.camila@ufabc.edu.br}
\emailAdd{y.villamizar@ufabc.edu.br}
\emailAdd{walten@hephy.oeaw.ac.at}

\abstract{\smo is a public tool for fast reinterpretation of LHC searches for new physics based on a large database of simplified model results. While previous versions were limited to models with a $\Ztwo$-type symmetry, such as R-parity conserving supersymmetry, version~3 can now handle arbitrary signal topologies. To this end, the tool was fully restructured and now relies on a graph-based description of simplified model topologies. 
In this work, we present the main conceptual changes and novel features of \smothree, together with the inclusion of new experimental searches for resonant production of spin-1 and spin-0 mediators with decays to quarks or to dark matter. 
Applying these results to a model containing two mediators, we discuss the interplay of resonance and missing energy searches, and the model's coverage by the currently available simplified model results.}

\begin{document} 
\maketitle
\flushbottom


\section{Introduction}

Searches for new physics at the LHC are sensitive to a far greater set of theories and parameter combinations than considered in the publications of the experimental collaborations. The reinterpretation of these searches, in order to fully understand their implications for new physics, is therefore of vital interest to the particle physics community. 

In this context, the aim of \smodels~\cite{Kraml:2013mwa,Ambrogi:2017neo,Ambrogi:2018ujg,Alguero:2021dig,MahdiAltakach:2023bdn} is to allow for the fast reinterpretation of simplified model results from searches for new physics at the LHC without the need of a dedicated Monte Carlo simulation. 
The basic working principle is that a full input model (consisting of particle content, masses, production cross-sections, and decay widths) of a given beyond the Standard Model (BSM) scenario is decomposed into simplified model components with their corresponding signal weights. 
These are then used to evaluate the constraints from a large database of 
experimental results. The outcome is by default presented in the form of so-called $r$-values, i.e.\ the ratio of the signal cross-section to its corresponding upper limit, but (global) likelihood analyses are also possible~\cite{MahdiAltakach:2023bdn,Altakach:2023tsd}. 
The speed and ease of use offered by \smodels are particularly relevant for model explorations and large scans. 
The approach requires that kinematic distributions of the tested signal and of the simplified model it is mapped onto are sufficiently similar such that the same cut acceptances apply (see \cite{Kraml:2013mwa,LHCRiF:2020xtr}).
Therefore it is complementary to simulation-based recasting~\cite{LHCRiF:2020xtr}, which may allow for
more precise reinterpretation
but is computationally much more expensive.
In addition, \smodels' applicability is limited to the simplified models included in the database and it is therefore relevant to have as large a database as possible.

So far, \smodels was based on the assumption that the simplified models followed a $\mathcal{Z}_2$ preserving structure, where BSM particles ($X,Y,...$) are always pair produced  
and decay as $X \to Y + \mbox{(any number of SM particles)}$. This limited the applicability of the tool to BSM scenarios with a conserved new parity; signal topologies like new $s$-channel resonances, 
associated production of BSM plus Standard Model (SM) particles, 
and final states consisting of only SM particles could not be treated.  

This limitation is overcome in version~3 of \smodels, presented in this paper. 
Concretely, version~3 generalises the applicability of the \smodels framework, so it can handle arbitrary simplified model topologies, without the need of an imposed $\mathcal{Z}_2$ symmetry. 
To achieve this, the tool was fully restructured and now relies on a graph-based description of simplified model topologies. 

In this work, we present the main conceptual changes to the \smodels tool and its API. Moreover, we demonstrate the physics impact of the new developments by means of a case study for the Two-Mediator Dark Matter (2MDM) model. This model extends the SM by a $\up$ gauge group and a scalar singlet $\phi$, which spontaneously breaks the new symmetry, giving rise to a massive gauge boson ($\zp$). Moreover, a Majorana singlet fermion $\chi$ is introduced as a dark matter (DM) candidate.  The model thus presents a variety of signatures, which provide a good showcase for the new features of \smodels v3.

The paper is organised as follows. Section~\ref{sec:graphs} presents the new, graph-based topology description of \smodels. This goes into some technical details regarding the inner workings of the tool, but is not directly relevant for the end-user. 
Changes to the user interface are explained in Section~\ref{sec:user}.  
The newly included experimental results are presented in Section~\ref{sec:database}. 
Section~\ref{sec:twoMDM} then presents the application to the 2MDM model. 
Here, we first explain the model and its signatures at the LHC (Sections~\ref{sec:model} and \ref{sec:signal}) and then discuss in detail the results obtained with \smodels (Section~\ref{sec:results}).  
We conclude in Section~\ref{sec:conclusions}.  
Two appendices complete the paper: Appendix~\ref{app:database} provides additional information about new results and technical improvements in the database, and Appendix~\ref{app:2mdm} gives more details on the 2MDM model.\\

\paragraph{Data management:} \smodels is published under the GPL v3.0 licence and 
available from \url{https://smodels.github.io}. 
A detailed description of the software, installation instructions, and more are given in the online manual \url{https://smodels.readthedocs.io}. 
The data and code used to produce the results of the phenomenological study
are published on Zenodo~\cite{smo3:dataset}, making the plots of Section~\ref{sec:results} fully reproducible.

\section{New Graph-Based Topology Description} \label{sec:graphs}

\subsection{Graph-based topologies}

The simplified model topologies (or ``SMS topologies'') used by \smodels are based on a few simplifying assumptions, which allow the tool to handle complex models with a minimal amount of input information. In particular, we assume that:
\begin{itemize}
    \item the details of the production mechanism can be ignored, such that the production is simply specified by the total cross-section and the new particles produced in the $p p$ collisions, and 
    \item all the decays of a BSM particle are described by its total decay width and the branching ratios in terms of on-shell decay products.
\end{itemize}

\begin{figure}[!t] \centering
    \includegraphics[width=0.7\textwidth]{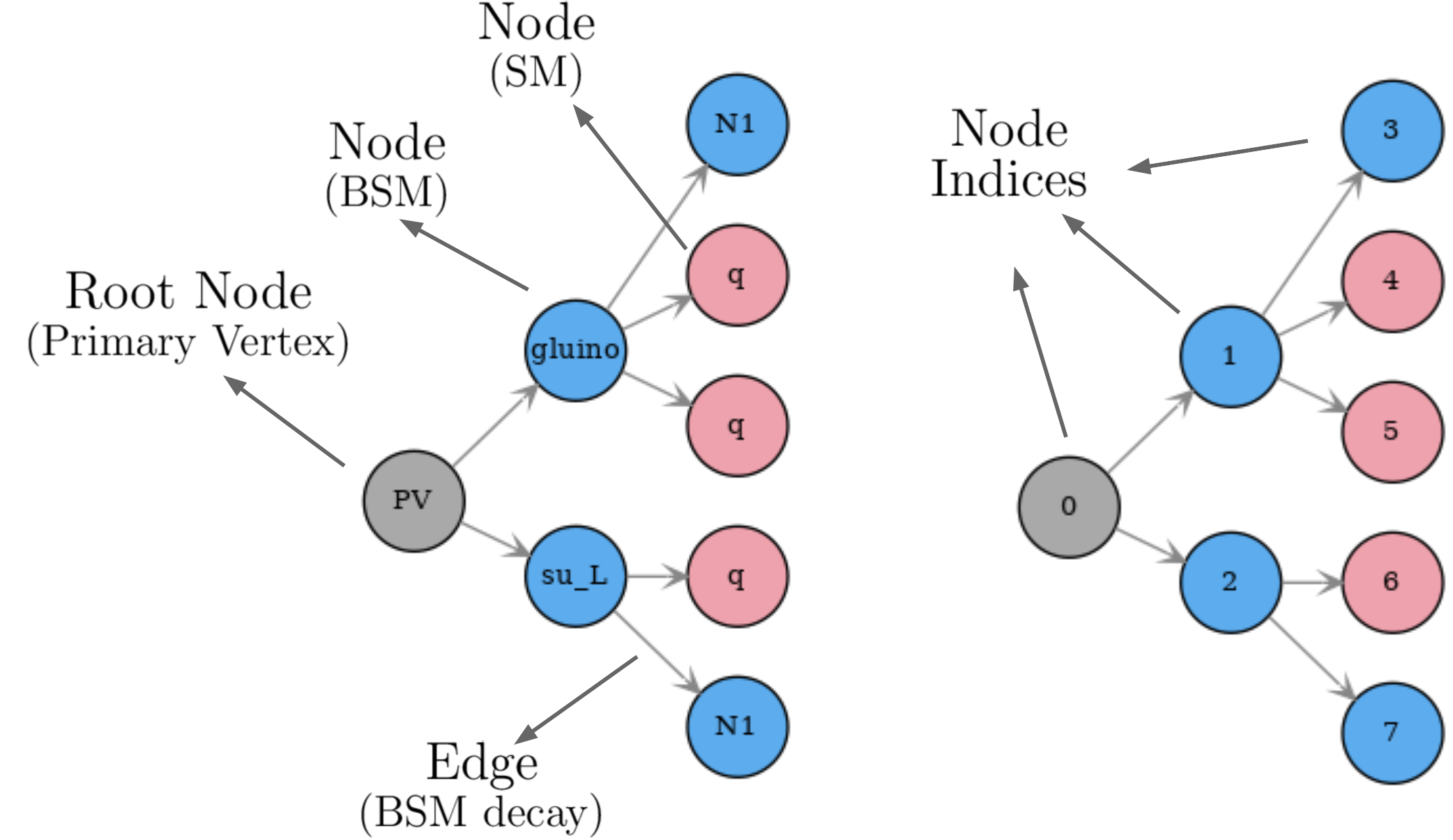}
    \caption{Graph representation of a simplified model topology and its elements: root node, SM and BSM nodes, edges and node indices.}
    \label{fig:smsScheme}
\end{figure}

Under these assumptions, the SMS topologies relevant for LHC searches can be described by means of graphs, concretely by directed rooted trees, as exemplified in Fig.~\ref{fig:smsScheme}. 
The root node represents the hard scattering collision ($p p \to \mbox{produced particles}$) and is labelled "PV" (primary vertex). All the particles appearing in the SMS topology correspond to graph nodes, while the decays are represented by edges connecting the parent particles to their daughters. 
The nodes appearing in the SMS topology are numbered by node indices for convenience, and hold all required information, i.e.\ quantum numbers, mass and total width, of the respective particle. 
Decays of SM states are not specified within the graph, since these are assumed to be given by the SM values.%
\footnote{It is worth pointing out that a similar graph description is also adopted by the HepMC3 library~\cite{Buckley:2019xhk} to describe Monte Carlo events. Within the \smodels framework, however, the ``events'' correspond only to the parton level process and are called SMS topologies.}

\begin{figure}[!t] \centering
    \begin{minipage}[c]{0.25\textwidth}
        \includegraphics[scale=0.4]{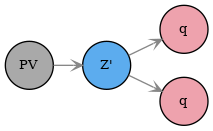}   
    \end{minipage} \hspace{15pt}
    \begin{minipage}[c]{0.25\textwidth}
        \includegraphics[scale=0.4]{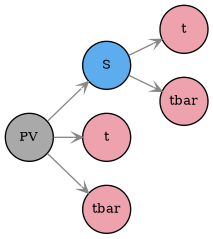}
    \end{minipage} \hspace{15pt}    
    \begin{minipage}[c]{0.25\textwidth}
        \includegraphics[scale=0.4]{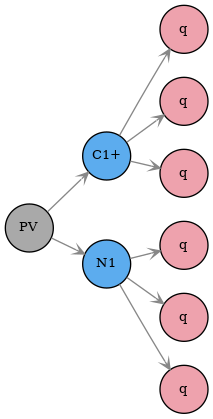}\\
        $\;$
    \end{minipage}
    \caption{Examples of new SMS topologies which can be handled by version~3.}
    \label{fig:nonz2_examples}
\end{figure}

This graph-based description is very flexible.
Specifically, it allows us 
to go beyond the two-branch structure (from pair-production of new particles followed by cascade decays), typical for models with a $\mathcal{Z}_2$-like symmetry.   
Examples of non-$\mathcal{Z}_2$ topologies which can now be handled in \smothree are shown in
Fig.~\ref{fig:nonz2_examples}.

The graph also holds some global information about the SMS topology, such as the total production cross-section, $\sigma$, and the product of all the branching ratios, $\prod_i {\rm BR}_i$, corresponding to the decays appearing in it. This information is used to compute the SMS \emph{weight}, $w_{\rm SMS}$, as  
\begin{equation}
    w_{\rm SMS} \equiv \sigma \times \prod_i {\rm BR}_i
\end{equation}
and in turn $r$-values, $r=w_{\rm SMS}/\sigma_{95}$, for upper limit (UL) type results.\footnote{The types of experimental results used in \smodels are explained in detail in the \href{https://smodels.readthedocs.io/en/latest/DatabaseDefinitions.html}{Database Definitions} section of the \href{https://smodels.readthedocs.io/}{online manual}.}
Here, $\sigma_{95}$ is the signal cross-section excluded at 95\% confidence level.
The SMS weight may also include an efficiency factor $\epsilon$ when used to compute the signal yields for experimental signal regions (SRs); in that case 
\begin{equation} 
    w_{\rm SMS} \equiv \epsilon\times \sigma \times \prod_i {\rm BR}_i 
\end{equation}
for each SR, corresponding to a fiducial cross-section. This is used with efficiency map (EM) type results and is particularly necessary for computing likelihoods~\cite{Ambrogi:2017neo,MahdiAltakach:2023bdn}.

\subsubsection*{String representation}

Although the graphical representation of SMS topologies shown in Fig.~\ref{fig:smsScheme} is extremely useful for visualising the topology, it is not always the most convenient. For instance, when describing the topologies in textual format, a one-line representation is needed. Within \smodels both formats are interchangeable, so a graph can also be represented in string format using a sequence of decay patterns of the type:
\begin{verbatim}
   X(i) > A(j),B(k),C(l)
\end{verbatim}
where $X$ represents a BSM particle which undergoes a three-body decay to (on-shell) particles $A$, $B$ and $C$. The indices $i,j,k,l$ refer to the node indices for the unstable particles in the SMS graph and are needed to avoid ambiguities. To give a concrete example, the SMS from Fig.~\ref{fig:smsScheme} is represented by the string:
\begin{verbatim}
   (PV > gluino(1),su_L(2)), (gluino(1) > N1(3),q(4),q(5)), (su_L(2) > q(6),N1(7))
\end{verbatim}
To make this simpler, by default only the indices of decaying BSM particles are shown in the \smodels output:
\begin{verbatim}
   (PV > gluino(1),su_L(2)), (gluino(1) > N1,q,q), (su_L(2) > q,N1)
\end{verbatim}
The string representation is also 
used when specifying the SMS topologies constrained by
experimental results in the \smothree database. 
Note, however, that the BSM particle names, like ``gluino'' or ``N1'', in this string representation are merely a convenient short-hand notation for generic BSM particles with the corresponding quantum numbers; they need not be SUSY particles.

\subsubsection*{Canonical name}

\begin{figure}[t]
    \centering
    \includegraphics[width=0.7\textwidth]{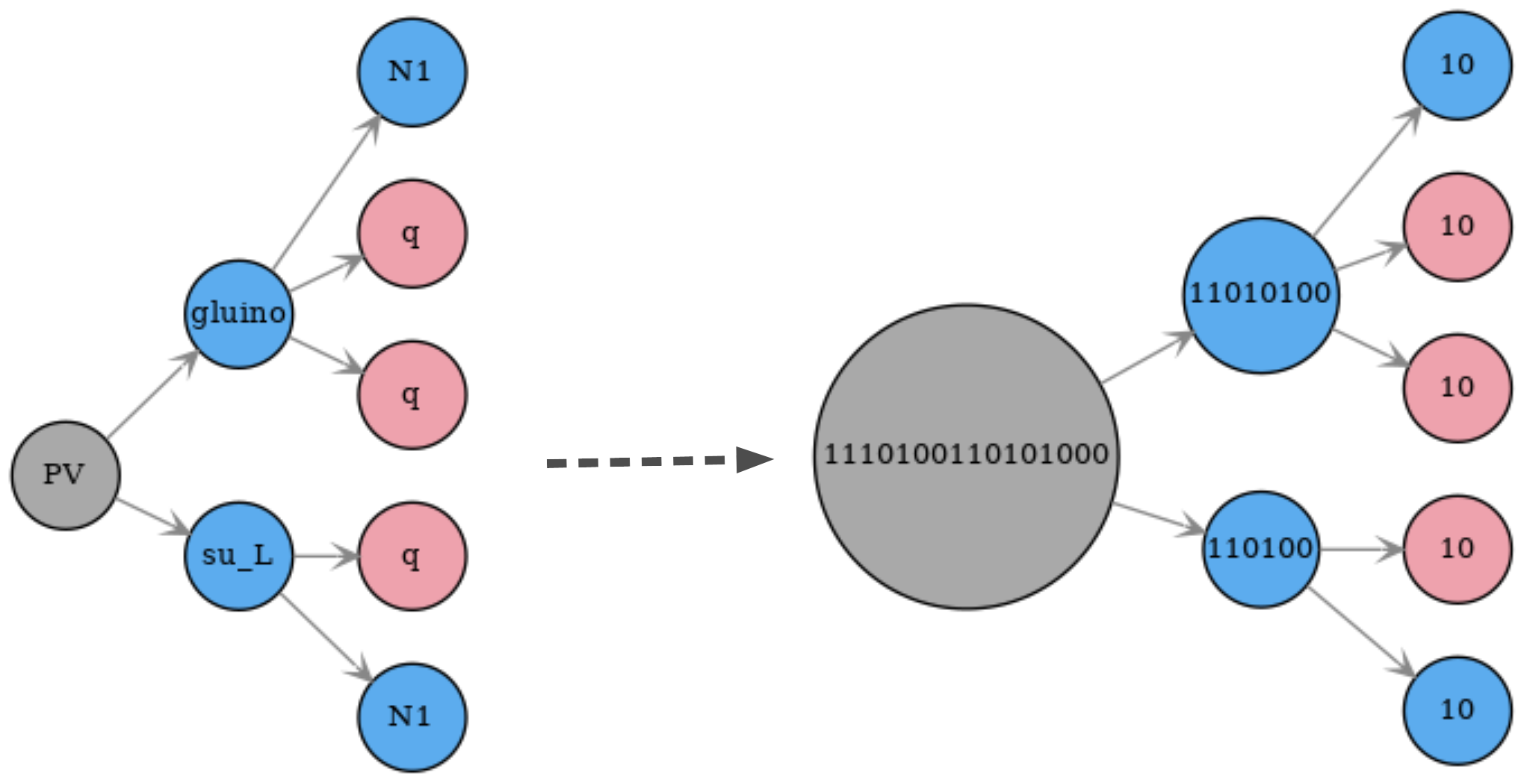}
    \caption{Example of how the canonical name is defined for each node. The SMS canonical name corresponds to the label of the primary vertex node.}
    \label{fig:canonfig}
\end{figure}

It is often useful (see discussion below in Section~\ref{sec:match}) to describe the \emph{structure} of an SMS topology without specifying its particle contents. 
This can be achieved using the canonical name (or canonical labelling) convention for rooted graphs, which assigns to each node a label according to the following rules:
\begin{itemize}
    \item each undecayed (final node) receives the label "10" 
    \item each decayed node receives the label "1<sorted labels of daughter nodes>0"
\end{itemize}
where "<sorted labels of daughter nodes>" is the joint string of the daughter nodes labels, sorted by size. Finally, the label associated with the "PV" node (root node) uniquely describes the graph structure. An example is shown in Fig.~\ref{fig:canonfig}.

\subsubsection*{Input Model and Database Graphs}

The graphs representing the input model are constructed using the quantum numbers, branching ratios and production cross-sections provided as input (see Section~\ref{sec:input} for more details) through the decomposition procedure~\cite{Alguero:2021dig}.
On the other hand, the graphs describing the SMS topologies constrained by the experimental searches in the \smo database are defined using the string representation discussed above. In this case a set of pre-defined BSM particles and their properties is used to map the particle strings to the objects with the correct particle properties. The text version of the database contains the file {\tt databaseParticles.py}, which lists all the available BSM particles used for describing the experimental results. This file can be readily modified by the user to include new particles, if needed. 

\subsection{SMS matching} \label{sec:match}

\begin{figure}[t]
    \centering
    \includegraphics[width=0.6\textwidth]{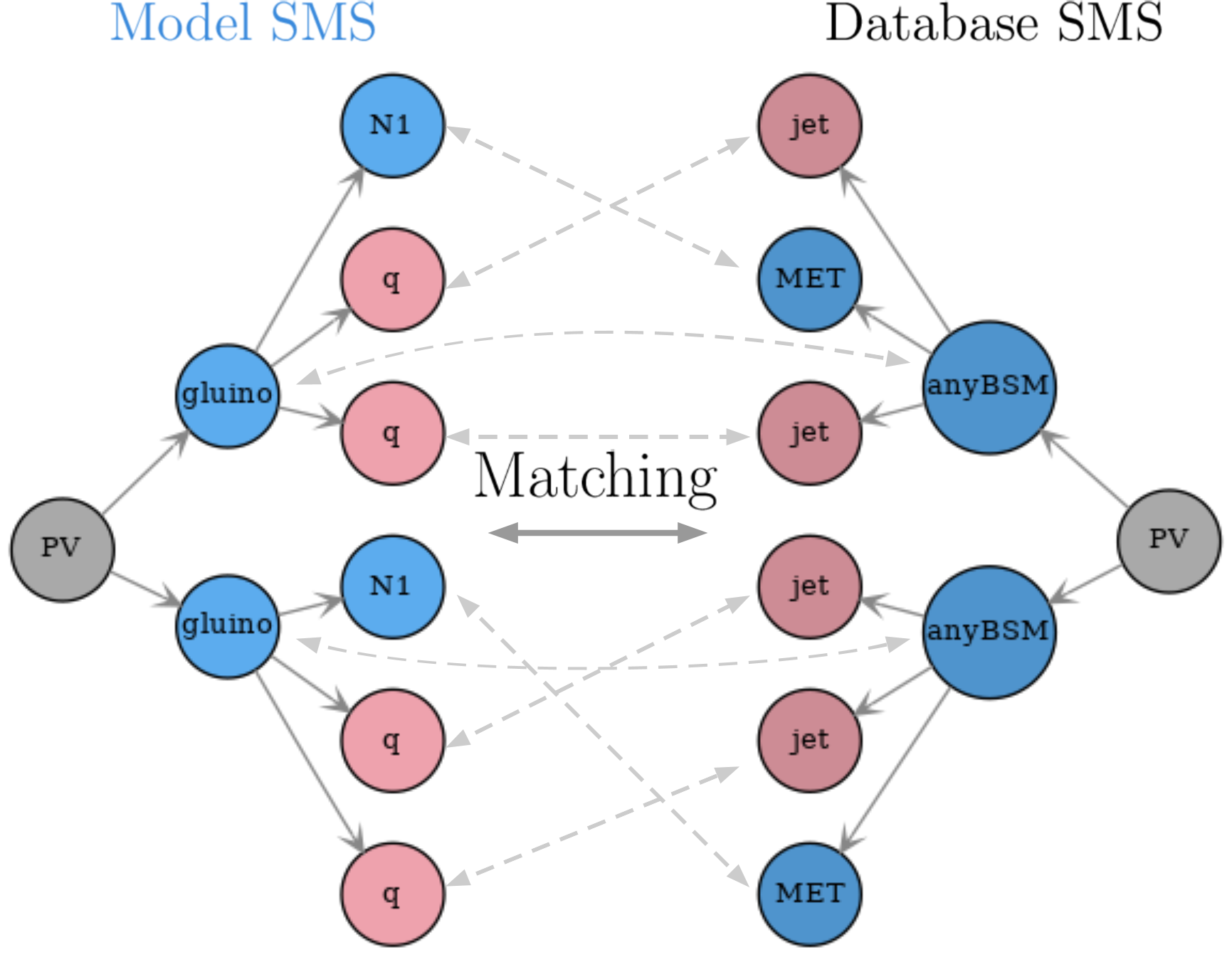}
    \caption{Schematic representation of the matching between SMS topologies generated by the decomposition (Model SMS) and the topologies found in the database (Database SMS).}
    \label{fig:tpA}
\end{figure}

Once all SMS topologies occurring in the input model have been determined, they need to be compared against those describing the experimental results in the \smodels database. This {\it matching}, schematically illustrated in Fig.~\ref{fig:tpA}, is a crucial step in the \smodels procedure in order to set constraints.
It had to be considerably modified with respect to the previous \smodels versions in order to account for the graph structure now used in version~3. 

\begin{figure}[t]
    \centering
    \includegraphics[width=0.6\textwidth]{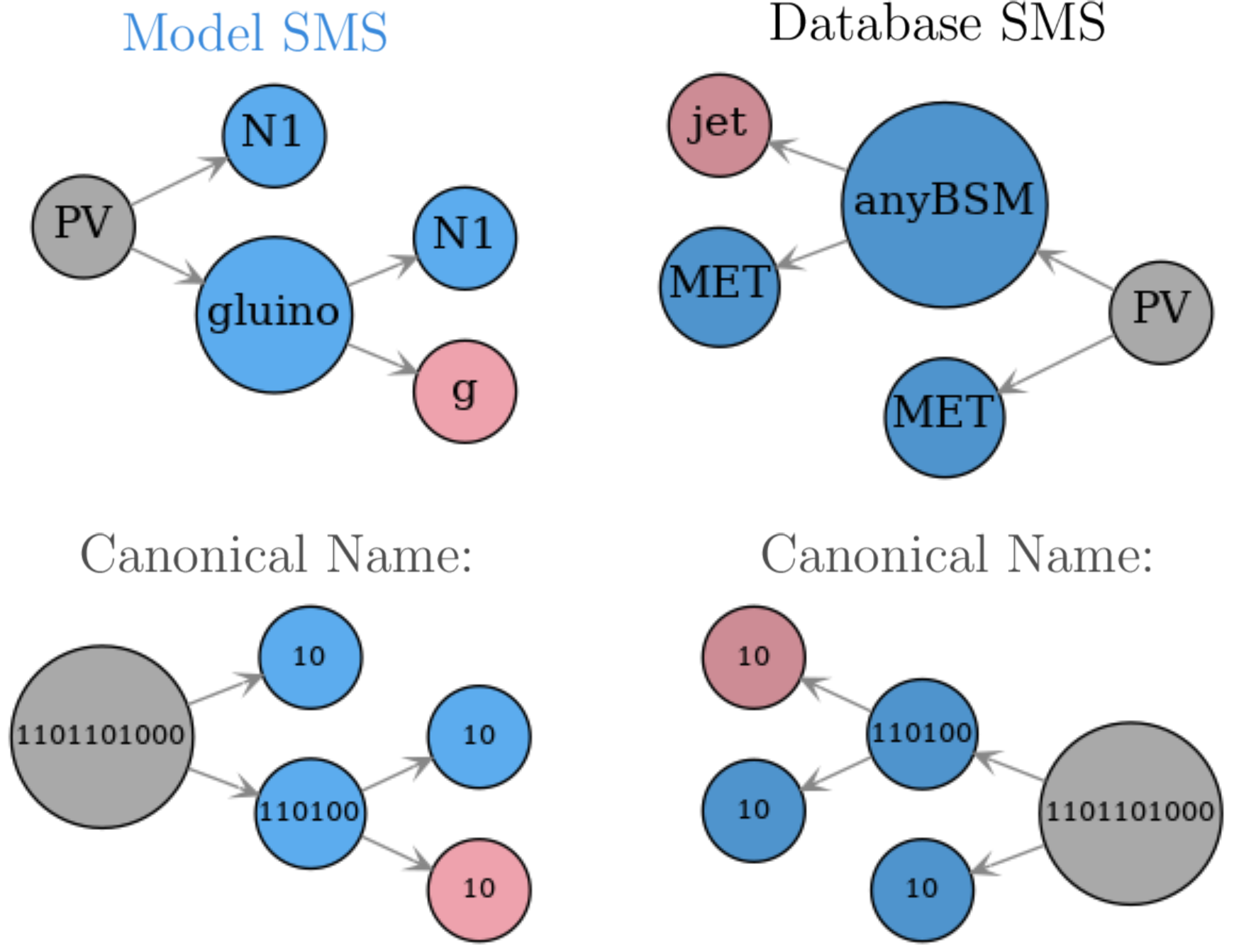}
    \caption{Example of two topologies to be matched and the respective canonical names for their nodes.}
    \label{fig:matchA}
\end{figure}

Two topologies are considered to match if {\it i)}~they have the same structure and {\it ii)}~the particles appearing in each topology have the same properties (electric charge, color representation, spin,...).\footnote{The comparison of the particle properties is done only for the properties which have been defined for both particles. For instance, in many cases the spin property is not defined for particles appearing in the database topologies, so this property will be ignored when comparing particles from the model topology to the ones from the database topology.}
In the language of graphs, any two nodes will be considered matched if:
\begin{enumerate}
    \item their canonical names are equal,
    \item their particle attributes match and
    \item their daughter nodes match {\it irrespective} of their ordering.
\end{enumerate}
With these criteria, the SMS topologies are traversed following a depth-first search, starting from the root nodes, until all nodes have been matched (if possible). To illustrate this procedure, let us consider the example of the model and database topologies shown in Fig.~\ref{fig:matchA}. In this example, the model SMS represents the associated production of a gluino and a stable neutralino (N1), with the gluino further decaying to a gluon (g) and a neutralino (N1). The database SMS shown in the figure is more inclusive and can be matched to any BSM particle (anyBSM) being produced in association with a second particle that leads to missing energy (MET). In addition, anyBSM is required to decay to a particle leading to MET and a jet, which corresponds to any light quark flavour or a gluon.

\begin{figure}[t] \centering
    \includegraphics[width=1\textwidth]{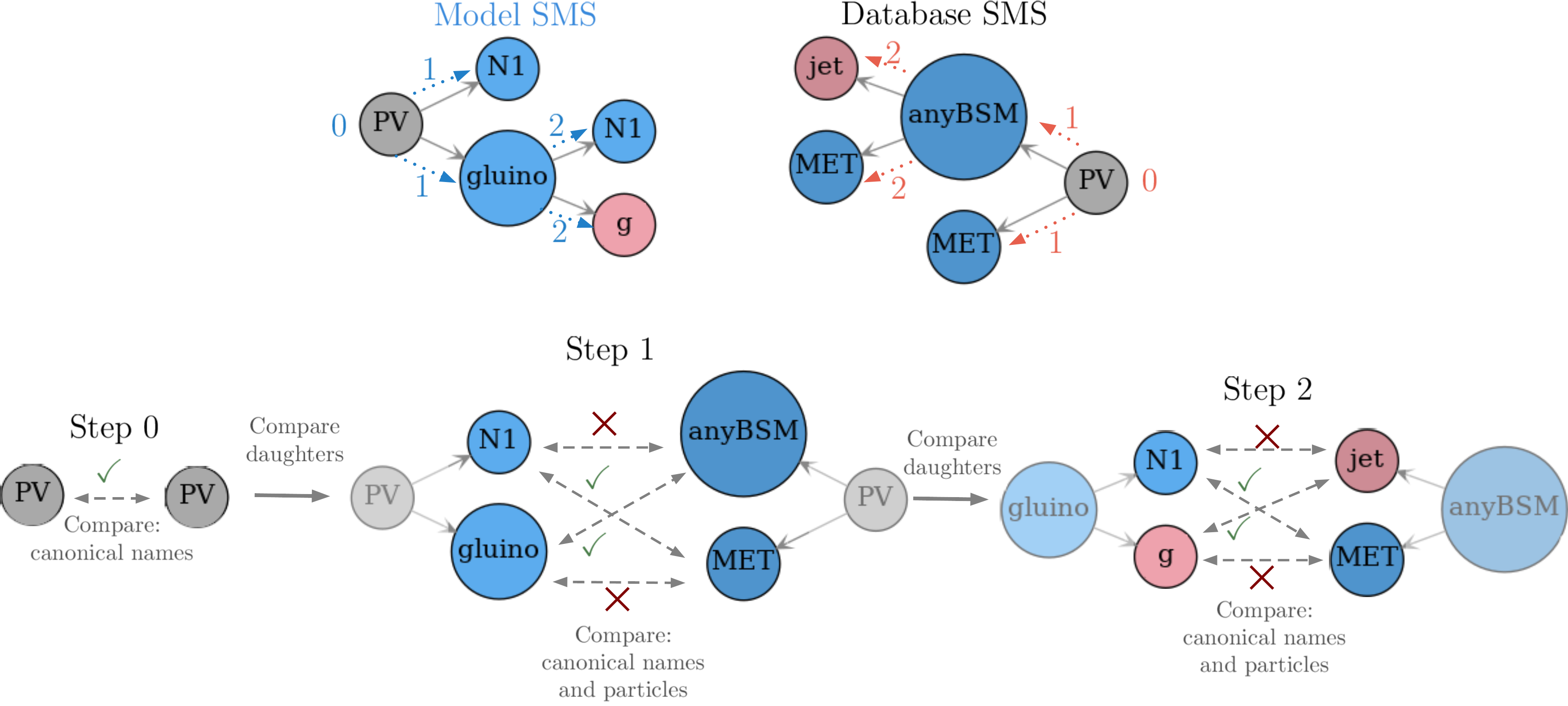}
    \caption{Step-by-step illustration of the matching procedure.}
    \label{fig:matchB}
\end{figure}

The matching procedure itself is illustrated in Fig.~\ref{fig:matchB}. It starts by comparing the root nodes, which in this example have the same canonical names (this enforces that both SMS have the same structure) and the same particle properties (which is always assumed as true for root nodes). This is indicated by {\it Step 0} in Fig.~\ref{fig:matchB}.
Hence criteria 1.\ and 2.\ for matching two nodes are satisfied. 
The next step consists in verifying if the third criterion is true, i.e.\ the root nodes daughters have to be compared {\it irrespective of their order}. In this example these are (gluino, N1) from the model SMS and (MET,\,anyBSM) from the database SMS. Once again we compare their canonical names and particle properties ({\it Step 1} in Fig.~\ref{fig:matchB}). At this stage, the comparison has to be made for any ordering of the daughter nodes:\footnote{In order to compare two sets of daughters (mapped to a bipartite graph) irrespective of their ordering, a maximal matching algorithm~\cite{enwiki:1221446268} is used. Note that in principle it is possible that the matching is not unique (i.e.\ $A \leftrightarrow a$, $B \leftrightarrow b$ and $A \leftrightarrow b$, $B \leftrightarrow a$)  
although we have not encountered this problem in practice.}
\begin{itemize}
    \item (gluino, N1) $\stackrel{?}{\simeq}$ (MET,anyBSM) or (N1, gluino) $\stackrel{?}{\simeq}$ (MET,anyBSM)
\end{itemize}
The label anyBSM can match {\it any} BSM particle, hence it matches N1. However, since their canonical names are different, Step~1 results in the following partial matches:
\begin{itemize}
    \item gluino $\leftrightarrow$ anyBSM, N1 $\leftrightarrow$ MET (Step 1)
\end{itemize}

\begin{figure}[t] \centering
    \includegraphics[width=0.66\textwidth]{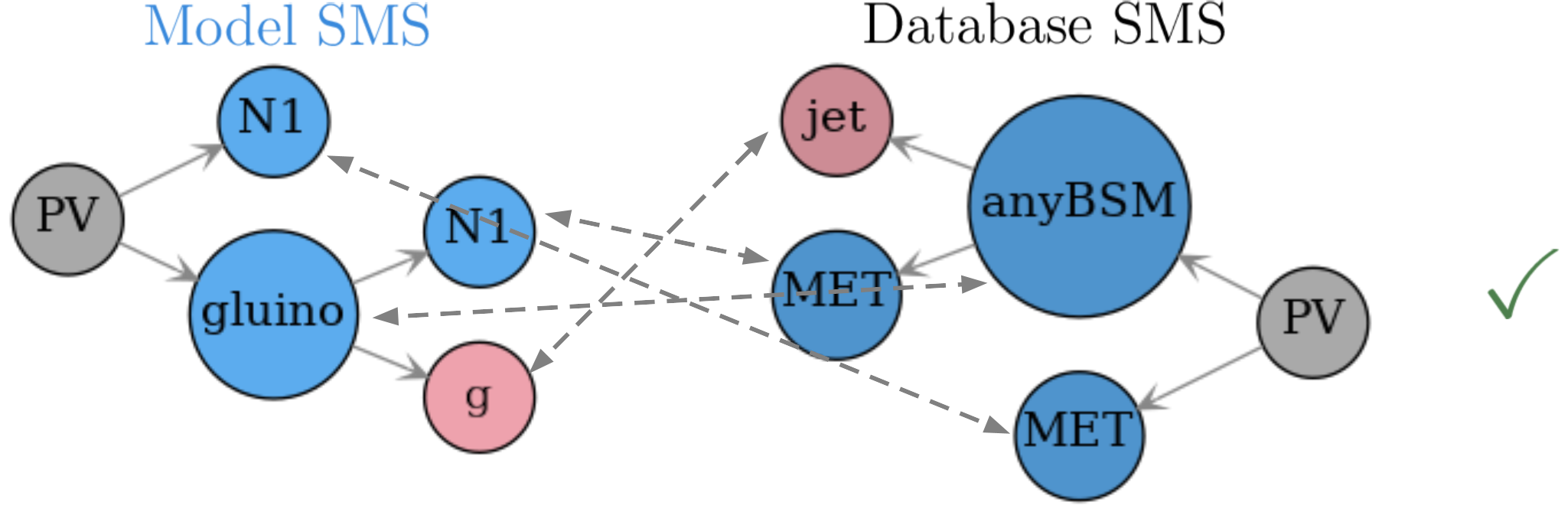}
    \caption{Result of the matching of the model and database topologies from Fig.~\ref{fig:matchA} and the identification between particles (nodes).}
    \label{fig:matchC}
\end{figure}

\noindent 
In order to fully match the gluino and anyBSM nodes,  their daughters must also be compared (Step 2). Since their daughters (g,\,N1) and (MET,\,jet) are final state nodes (undecayed) the comparison procedure stops at this level with the result
\begin{itemize}
    \item g $\leftrightarrow$ jet, N1 $\leftrightarrow$ MET (Step 2)
\end{itemize}
This means a full match for the (gluino,\,N1) and (anyBSM,\,MET) pairs, which then means that the root nodes fully match. Consequently, the model and database topologies match (see Fig.~\ref{fig:matchC}) with the identifications: gluino $\leftrightarrow$ anyBSM, N1 $\leftrightarrow$ MET, N1 $\leftrightarrow$ MET and g $\leftrightarrow$ jet.
This procedure can be applied to any pair of SMS topologies.

\section{Changes in the User Interface}\label{sec:user}

Although the structural changes described in Section~\ref{sec:graphs} allow \smothree to deal with arbitrary signal topologies, the tool's API remains very similar to previous versions. Nonetheless, version 3 introduces small changes in the \smodels input and output, which are described in this section.

\subsection{Changes regarding the input model and parameter card} \label{sec:input}

The input model defines the BSM states and their quantum numbers. It can be
provided either as a python module or as an SLHA file containing QNUMBERS blocks.\footnote{A path to the user's own model file (in either format) can be specified in the {\it [particles]} section of the parameter card (\code{parameters.ini}).}
In previous versions, the definition of the BSM particles and their quantum numbers had to include the \code{Z2parity} quantum number, since a $\mathcal{Z}_2$ parity was assumed to be conserved. This is no longer needed in version~3 and, if defined, this quantum number is ignored.
In the case of using a python module as input,  BSM states are defined using the following syntax:
\begin{verbatim}
bsm = Particle(isSM=<True/False>, label=<particle label>, pdg=<pdg number>, 
      eCharge=<electric charge>, 
      colordim=<color representation>, spin=<spin>)
\end{verbatim}
For instance, the left-handed down squark in the MSSM would be defined as:
\begin{verbatim}
sdl = Particle(isSM=False, label='sd_L', pdg=1000001, 
               eCharge=-1./3, colordim=3, spin=0)
\end{verbatim}
More details are provided in the \href{https://smodels.readthedocs.io/en/stable/BasicInput.html}{Basic Input} section of the \href{https://smodels.readthedocs.io}{online manual}.

Although not new, it is important to note here, that the properties of SM particles (masses, branching ratios, etc.) are fixed in \smodels and cannot be modified through the input model. This also applies to the SM-like Higgs boson, which is assumed to have a mass of 125~GeV and SM branching ratios, and is associated with the PDG code 25. Therefore, if the BSM model has an enlarged Higgs sector, a PDG code 25 must be assigned \emph{only} to a 125~GeV Higgs with SM-like decay branching ratios; for non-SM-like scalars other PDG codes should be used. This is important for correctly matching experimental results involving a SM-like Higgs boson for which the analysis assumed SM branching ratios.

Regarding the parameter card (\code{parameters.ini}), which lets the user control most of the \smodels behaviour, two new options were added in version~3:
\begin{enumerate}
    \item {\tt ignorePromptQNumbers}: list of particle attributes to be ignored for promptly decaying BSM particles. Since many experimental searches are not sensitive to the detailed properties of particles with prompt decays, \smodels has the option to    
    ignore certain quantum numbers of these particles. For instance, if \code{ignorePromptQNumbers = "spin,eCharge,colordim"}, the spin, electric charge and color properties of promptly decaying particles will be ignored. This can greatly reduce the running time, but must be used with caution. If  
    \code{ignorePromptQNumbers} is not defined, all quantum numbers will be kept. 
    \item {\tt outputFormat}: type of output format in which the output should be written. As discussed in Section~\ref{sec:graphs}, the SMS topologies 
    can be cast in a string representation describing the production and decays of BSM particles. This is the default format adopted in version 3, but setting \code{outputFormat = version2} will write the output using the bracket notation~\cite{Kraml:2013mwa} adopted in previous versions. For SMS topologies, which cannot be written in bracket notation, the version~3 output will be used.
\end{enumerate}
A detailed description of all options and parameters can be found in the \href{https://smodels.readthedocs.io}{online manual}.

\subsection{Changes in the output}

The output of version~3 follows the same structure of previous versions with the exception that the old bracket notation~\cite{Kraml:2013mwa} has been replaced by
the string representation described in Section~\ref{sec:graphs}. It is also possible to convert to the bracket notation format (if the topologies obey a $\mathcal{Z}_2$ symmetry) using the \code{outputFormat = version2} option described above. A detailed description of all the available output types can be found in the \href{https://smodels.readthedocs.io}{online manual}.
For illustration, we show an excerpt of the stdout output
using the old (version 2) and the new (version 3) formats.

\clearpage

\begin{itemize}
    \item Version 2: 
\begin{verbatim}
Element ID: 1
Particles in element: [[[higgs]], [[W-]]]
Final states in element: [N1, N1~]
The element masses are 
Branch 0: [2.69E+02 [GeV],1.29E+02 [GeV]]
Branch 1: [2.69E+02 [GeV],1.29E+02 [GeV]]

The element PIDs are 
PIDs: [1000023,1000022]
PIDs: [1000024,1000022]
The element weights are: 
  Sqrts: 1.30E+01 [TeV], Weight:3.92E-01 [pb]
  Sqrts: 8.00E+00 [TeV], Weight:1.74E-01 [pb]
\end{verbatim}

\item Version 3:
\begin{verbatim}
SMS ID: 1
SMS: (PV > N2(1),C1-(2)), (N2(1) > N1,higgs), (C1-(2) > N1~,W-)
Masses: [(N2, 2.69E+02 [GeV]), (C1-, 2.69E+02 [GeV]), (N1, 1.29E+02 [GeV]), 
         (N1~, 1.29E+02 [GeV])]
Cross-Sections:
  Sqrts: 1.30E+01 [TeV], Weight:3.92E-01 [pb]
  Sqrts: 8.00E+00 [TeV], Weight:1.74E-01 [pb]
\end{verbatim}
\end{itemize}

As shown by the example above, the new format relying on the string representation is more compact and informative. Information about the particle masses is displayed as a list of tuples, so it is clear which BSM particles the masses refer to. 
In addition to the default (textual) formats available in \smodels, it is also possible to obtain a graphical representation of the SMS topologies 
using \smodels as a python library. An \href{https://smodels.readthedocs.io/en/latest/drawingGraphs.html}{example} of how to do this is provided in the \href{https://smodels.readthedocs.io/en/latest/Examples.html}{How To's} section of the \href{https://smodels.readthedocs.io/}{online manual}.

Finally, \smodels can also provide information about the SMS topologies present in the input model, but not constrained by any experimental result in the \smodels database. Within the \smodels framework these topologies are called {\it missing topologies}. Since, depending on the model, the number of missing topologies can be too large to be displayed in the output, they are grouped according to their final state signatures. This gives an idea of how much signal cross-section is being ``missed'' in distinct channels.
The simplification of missing topologies has also been generalised in version 3, so it can be applied to arbitrary topologies. It corresponds to the following steps:
\begin{enumerate}
    \item the topology is reduced to the primary BSM particles and their final states, i.e.\ all information about the intermediate BSM states is removed;
    \item the final state particles are replaced by inclusive ones (e.g., $e^\pm,\mu^\pm \to l$).
\end{enumerate}

\begin{figure}[t!]
    \centering
    \includegraphics[width=0.7\textwidth]{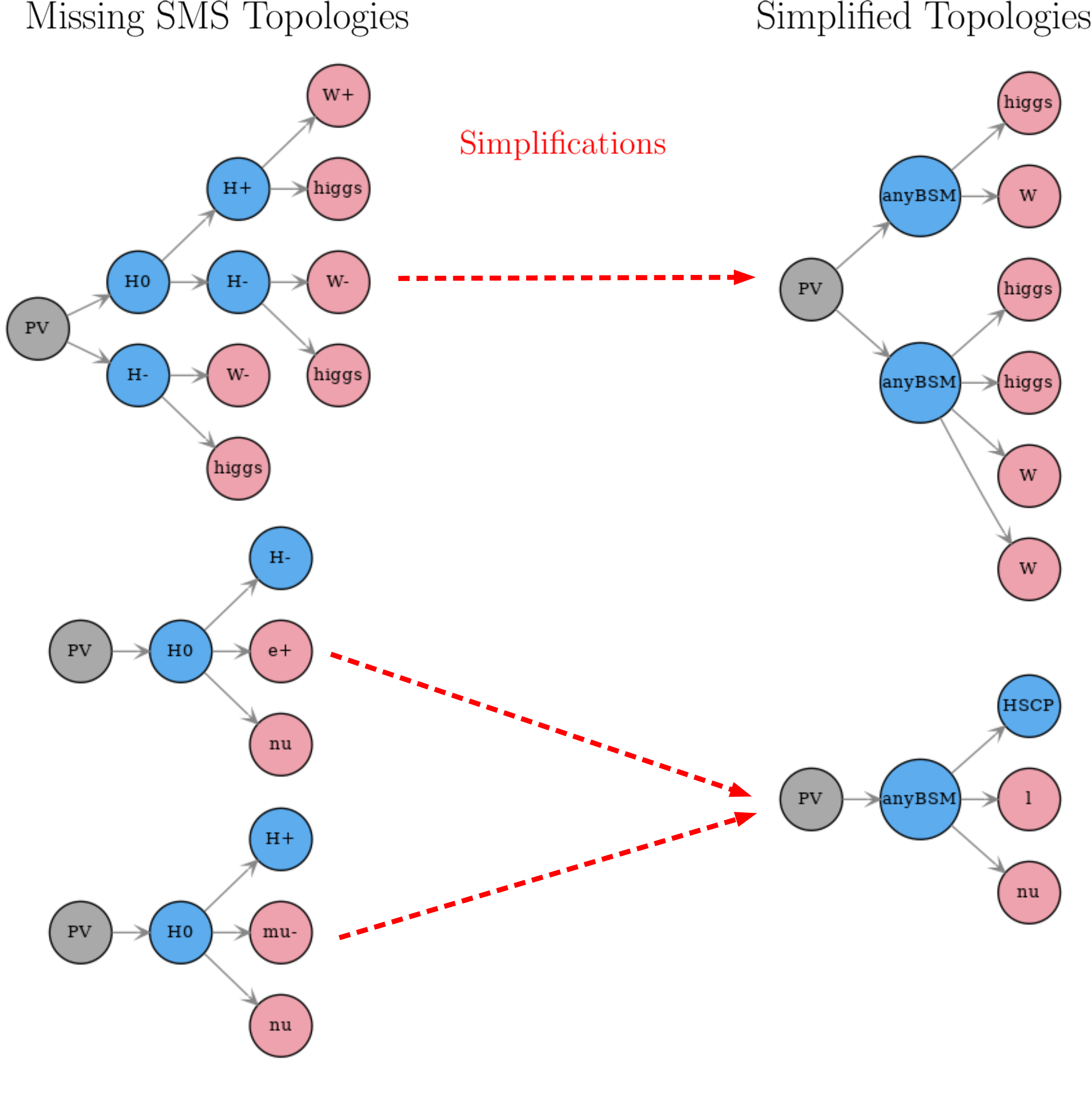}
    \caption{Schematic representation of how the missing topologies are simplified and combined in version 3.}
    \label{fig:smsCoverage}
\end{figure}

\noindent
Finally, after the above simplifications, identical missing (simplified) topologies are combined, i.e. their weights are added.
This procedure is illustrated in Fig.~\ref{fig:smsCoverage}. 
The simplified topologies are then represented in the output by their final states only, with all contributing cross-sections added up.
When printing the missing topologies using the string representation, only the set of final states is shown, grouped in parenthesis by their primary BSM state, e.g. {\tt (PV > X,Y), (X > a,b), (Y > A,B,C)} is compressed to {\tt PV > (a,b), (A,B,C)}.
A concrete example in summary-type output format is:
\begin{verbatim}
missing topologies with prompt decays with the highest cross-sections (up to 10):
Sqrts (TeV)   Weight (fb)               SMS description
 13.0          9.648E+01    #            PV > (higgs,W), (higgs,higgs,W,W) 
 13.0          7.018E+01    #            PV > (HSCP,l,nu) 
 13.0          4.897E+01    #            PV > (W,jet,MET), (W,b,t,MET) 
 ...
\end{verbatim}

\section{New Results in the Database}\label{sec:database}

\begin{figure}[t]\centering
    \includegraphics[width=0.25\textwidth] {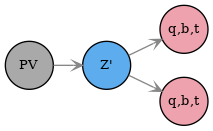}
    \hspace{20pt}
    \includegraphics[width=0.25\textwidth]{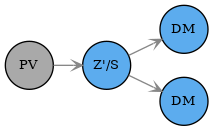}
    \caption{Resonant topologies covered by the new results in the database. The diagram on the right corresponds to the resonant production of a scalar ($S$) or a spin-1 boson ($\zp$) decaying to DM.}
    \label{fig:zprimeGraphs}
\end{figure}

The graph-based topology description
allows \smothree to include a number of experimental results that could not be considered in previous versions.
To illustrate this capability, we included in the database 10 new experimental results which focus on searches for spin-1 resonances ($Z'$) decaying to jets ($u,d,s \mbox{ or }c$ quarks), $b$-jets, top quarks, or DM as shown in Fig.~\ref{fig:zprimeGraphs}.  Moreover, we included results for the resonant production of a scalar ($S$), which decays to a pair of DM particles, as it is often considered in mono-X searches. The main difference between the scalar and the spin-1 production is that the former is assumed to proceed through a loop-induced gluon fusion diagram, while the latter occurs at tree-level in $q\bar{q}$ collision. 
Since these different production modes lead to different kinematic distributions, spin-0 and spin-1 resonances are treated as separate SMS.\footnote{We have verified that vector and axial-vector mediators have very similar kinematic distributions and therefore 
need not to be distinguished. The same is also valid for scalar and pseudo-scalar mediators.}
\cref{tab:database} gives a summary of these new analyses, the SMS topologies covered by each of them, and the type of information available: cross-section upper limit maps or efficiency maps.

\begin{table}[t!]
  \centering
  \caption{\label{tab:database} Summary of the new searches for non-\Ztwo\ topologies included in the \smothree database~\cite{smo3:database}. The column `SMS Topology' denotes the topologies constrained by the search, while the column `Type' specifies the type of result: upper limit (UL) or efficiency map (EM). The EM-type results were obtained through recasting (see text).\\}
  \rowcolors{2}{gray!10}{white}
  \hspace*{-6mm}
  \begin{tabular}{llrcc}
    \toprule
ID   
&  Signature \hphantom{1cm}  & Luminosity & SMS Topology & Type \\
\toprule 
\rowcolor{dukeblue!10!}  
\multicolumn{5}{c}{Run 2 - 13 TeV} \\
\href{https://atlas.web.cern.ch/Atlas/GROUPS/PHYSICS/PAPERS/EXOT-2019-03/ }{ATLAS-EXOT-2019-03} \cite{ATLAS:2019fgd}  & Dijet resonance &139 fb$^{-1}$ &$p p \to \zp \to j j,b\bar{b}$ & UL \\   
\href{https://atlas.web.cern.ch/Atlas/GROUPS/PHYSICS/PAPERS/EXOT-2018-48/}{ATLAS-EXOT-2018-48}
\cite{ATLAS:2020lks}    &$t\bar{t}$ resonance  &139 fb$^{-1}$ &$p p \to \zp \to t\overline{t}$ & UL\\
\href{https://cms-results.web.cern.ch/cms-results/public-results/publications/EXO-19-012/}{CMS-EXO-19-012}
\cite{CMS:2019gwf} & Dijet resonance &137 fb$^{-1}$ &$p p \to \zp \to j j,b\bar{b}$ & UL\\
\href{https://cms-results.web.cern.ch/cms-results/public-results/publications/EXO-20-008/}{CMS-EXO-20-008}
\cite{CMS:2022eud}      &$b$-jet resonance  &138 fb$^{-1}$ &$p p \to \zp \to b\overline{b}$ & UL\\
\href{https://cms-results.web.cern.ch/cms-results/public-results/publications/EXO-20-004/}{CMS-EXO-20-004}
\cite{CMS:2021far}      & Monojet &137 fb$^{-1}$ &$p p \to \zp,S \to \chi \chi$ & EM\\  
\href{https://atlas.web.cern.ch/Atlas/GROUPS/PHYSICS/PAPERS/EXOT-2018-06/}{ATLAS-EXOT-2018-06}  \cite{ATLAS:2021kxv} & Monojet & 139 fb$^{-1}$ &$p p \to \zp \to \chi \chi$ & UL\\ 
\href{https://atlas.web.cern.ch/Atlas/GROUPS/PHYSICS/PAPERS/SUSY-2018-22/}{ATLAS-SUSY-2018-22}  \cite{ATLAS:2020syg} & Multi-jet plus $\etmiss$ & 139 fb$^{-1}$ &$p p \to \zp \to \chi \chi$ & EM\\ 
\href{https://atlas.web.cern.ch/Atlas/GROUPS/PHYSICS/PAPERS/SUSY-2018-13/}{ATLAS-SUSY-2018-13}  \cite{ATLAS:2023oti} & Displaced jets & 139 fb$^{-1}$ &$p p \to \tilde\chi \tilde\chi \to jjj,jjj;...$  & EM\\ 
\rowcolor{dukeblue!10!}
\multicolumn{5}{c}{Run 1 - 8 TeV} \\
\href{https://cms-results.web.cern.ch/cms-results/public-results/publications/EXO-16-057}{CMS-EXO-16-057}
\cite{CMS:2018kcg}      &$b$-jet resonance  &19.7 fb$^{-1}$ &$p p \to \zp \to b\overline{b}$  & UL\\   
\href{https://cms-results.web.cern.ch/cms-results/public-results/publications/EXO-12-059}{CMS-EXO-12-059}
 \cite{CMS:2015rkn}   &Dijet resonance &19.7 fb$^{-1}$ &$p p \to \zp \to j j$ & UL\\ 
\href{https://atlas.web.cern.ch/Atlas/GROUPS/PHYSICS/PAPERS/EXOT-2013-11}{ATLAS-EXOT-2013-11} \cite{ATLAS:2014ktg} & Dijet resonance& 20.3 fb$^{-1}$ &$p p \to \zp \to  j j$  & UL\\ 
\bottomrule
\end{tabular}
\vspace*{2mm}
\end{table}

Two comments are in order here. 
First, the constraints on $\zp \to q\bar{q},\chi\chi$ may depend on the $\zp$ mass and width. For $\etmiss$ searches the dependence on $\Gamma_{\zp}$ is expected to be mild, since at truth level we have $\etmiss = p_T(\zp)$. Resonance searches~\cite{ATLAS:2019fgd,ATLAS:2020lks,CMS:2019gwf,CMS:2022eud,CMS:2018kcg,CMS:2015rkn,ATLAS:2014ktg}, however, rely on invariant mass measurements, which are directly affected by the width. Hence it becomes relevant to include the width dependence when considering resonance searches results.
The CMS-EXO-19-012~\cite{CMS:2019gwf} analysis provides limits for $\Gamma_{\zp}/m_{\zp}$ between 1\% and 55\% for $m_{\zp} > 1.8$~TeV, so this analysis can be used to constrain models with heavy and broad resonances. Unfortunately, the other resonance searches considered do not provide information on the $\zp$ width dependence; they can hence safely be applied only within the narrow width approximation (NWA),
which is taken to be valid for $\Gamma_{\zp}/m_{\zp} < 1\%$.\footnote{It is worth noting that ATLAS-EXOT-2019-03~\cite{ATLAS:2019fgd} provides width-dependent limits for a Gaussian model. However, for broad and heavy resonances it is not clear how these limits can be applied to a Breit-Wigner signal, hence these are not included in the database.}

Second, in addition to visible decays of resonances, decays to a pair of DM particles can be constrained by a variety of $\etmiss$ searches. Here, we  consider the monojet searches ATLAS-EXOT-2018-06~\cite{ATLAS:2021kxv} and 
CMS-EXO-20-004~\cite{CMS:2021far}, and the multi-jet plus $\etmiss$ search ATLAS-SUSY-2018-22~\cite{ATLAS:2020syg}.\footnote{Note that the monojet searches require one hard jet but allow for additional jets; the multi-jet searches require at least two jets.}
For the ATLAS monojet search, only the observed cross-section upper limits for $\zp\to \chi\chi$ are included in the database. For the CMS monojet and the ATLAS multijet searches, on the other hand, 
we produced dedicated efficiency maps for $p p \to \zp,S \to \chi\chi$ (+jets) by means of public recast codes. Concretely, for ATLAS-SUSY-2018-22 we used its implementation in CheckMATE2~\cite{Dercks:2016npn,Lara:2022new}, while for CMS-EXO-20-004 we used our own stand-alone code published on Zenodo~\cite{zenodoMonoJet}. 
The EMs were generated as a function of the mediator ($\zp$ or $S$) and DM ($\chi$) masses for the on-shell region, $m_{\zp,S} > 2 m_{\chi}$. 
As discussed above, for $\etmiss$ searches we expect a milder width-dependence~\cite{Chala:2015ama} and the efficiencies were computed within the NWA.
The implementation does not  enforce any constraint on $\Gamma/m$ and it is left to the users' judgement how large a $\Gamma/m$ they deem acceptable.
In the phenomenological analysis in Section~\ref{sec:twoMDM}, we will allow widths up to $\Gamma_{\zp}/m_{\zp} \simeq 5\%$ when computing limits using MET searches.

Statistical models in the form of a covariance matrix~\cite{CMS:SL} for CMS-EXO-20-004 and of a pyhf~\cite{Heinrich:2021gyp,pyhf} likelihood for ATLAS-SUSY-2018-22 are publicly available, allowing for a combination of the SRs within each individual analysis.

\begin{figure}[t!]
    \centering
    \begin{minipage}{.3\textwidth}
    \includegraphics[scale=0.35]{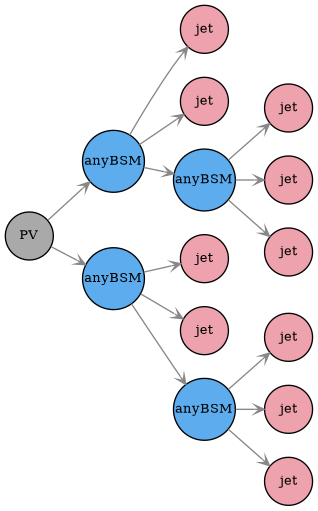}    
    \end{minipage}
     \hspace{5pt}
    \begin{minipage}{.3\textwidth}
    \includegraphics[scale=0.35]{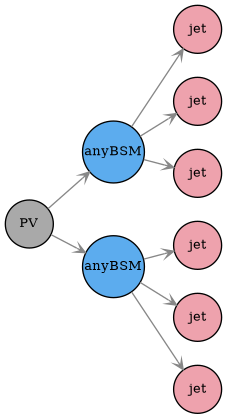}    
    \end{minipage}
    \caption{RPV topologies covered by the ATLAS-SUSY-2018-13~\cite{ATLAS:2023oti} analysis newly included in the database.}
    \label{fig:rpvGraphs}
\end{figure}

Besides the above resonance and $\etmiss$ searches, we have also included for the first time searches for R-parity violating (RPV) supersymmetry, 
in particular ATLAS-SUSY-2018-13~\cite{ATLAS:2023oti}, which looks for BSM particles with displaced decays to jets, see~\cref{tab:database}. 
In this case, we implemented EMs for the topologies shown in Fig.~\ref{fig:rpvGraphs} as a function of the relevant BSM masses and widths.
These EMs were computed using the recasting code \cite{llprepo:atlasrpv} available in the \href{https://github.com/llprecasting/recastingCodes}{LLP recasting code GitHub repository}.
Additional new results included in the database are presented in Appendix~\ref{app:newresults}.

\section{Physics Application: 2MDM Model}\label{sec:twoMDM}

In this section, we demonstrate the physics impact of \smothree by means of 
a case study for the Two-Mediator Dark Matter (2MDM) model~\cite{Duerr:2016tmh,Argyropoulos:2021sav,Duerr:2017uap}.
While this model closely resembles the simplified DM models considered by the experimental searches, it has three interesting features: i) it contains two DM mediators ($S$ and $\zp$), ii) it explains the $\zp$ and DM masses through the symmetry breaking induced by the new scalar $S$ 
and iii) it allows for a wider range of mediator and DM masses consistent with the observed DM relic abundance.
We point out, however, that the model is not anomaly free~\cite{FileviezPerez:2015mlm,Ellis:2017tkh,Butterworth:2024eyr}, and requires the introduction of other new particles, which we assume to have masses of several TeV or above. 
Sections~\ref{sec:model}--\ref{sec:results} present the main features of the model, its signals at the LHC and the results obtained with \smothree; additional details on the model are given in Appendix~\ref{app:2mdm}.

\subsection{Model}\label{sec:model}
The 2MDM model extends the SM by promoting the global baryon number symmetry to a gauge symmetry, denoted as $\up$.
As a result, only the SM quarks are charged under this symmetry and their charges are universal, i.e.\ they are the same for the left- and right-handed quarks of all three generations.
The $\up$ implies a new gauge boson ($\zp$). In addition,
a complex scalar ($\phi$) and a Majorana fermion ($\chi$) are introduced. Both are singlets under the SM gauge symmetries, but charged under the $\up$. In particular,  
$\chi$ transforms as $\chi \to e^{i g_{\zp} q_\chi \alpha(x) \gamma^5} \chi$.
The charges of the SM and BSM fields under the new $\up$ symmetry are presented in \cref{tab:u1charges}. 

\begin{table}[h!]
  \centering
   \caption{\label{tab:u1charges} Field content and $\up$ charges of the 2MDM model. The $q_L$, $u_R$, $d_R$, $l_L$ and $l_R$ are the left-handed quark doublet, the right-handed up-type and down-type quarks, the left-handed lepton doublet and the right-handed lepton, respectively. The Higgs doublet is denoted by $H$, while the BSM scalar singlet and Majorana fermion are denoted as  $\phi$ and $\chi$, respectively.}
  \rowcolors{2}{gray!10}{white}
  \vspace{0.2cm}
  \begin{tabular}{c||cccccccc}
      \toprule
$\psi$ & $q_{L}$ & $u_{R}$ & $d_{R}$ & $l_{L}$ & $l_{R}$ & $H$ & $\phi$ & $\chi$ \\
\toprule 
\rowcolor{dukeblue!10!} 
$\up$ & $q_q$ & $q_q$ & $q_q$ & 0 & 0 & 0  & $q_{\phi}$ & $q_{\chi}$\\
       \bottomrule
    \end{tabular}
\end{table}

Despite its minimality, the 2MDM allows one to explain both the $\zp$ and $\chi$ masses through the $\up$ spontaneous symmetry breaking (SSB) triggered by the vacuum expectation value (vev) of the new scalar $\phi$. 
Moreover, assuming $m_\chi < M_\zp,M_S$, the $\chi$ is stable thus providing a DM candidate. 
In this work, however, we focus purely on the LHC constraints.  We will neither enforce the model to generate the observed DM abundance, nor to satisfy limits from direct detection experiments, since these constraints can be evaded, e.g., by a non-standard cosmological evolution, multi-component DM scenarios and/or the introduction of new BSM states at higher scales. 

The Lagrangian of the 2MDM model is given by: 
\begin{equation}
    \mathcal{L} = \mathcal{L}_{\text{SM}}+ \mathcal{L}_{\zp} + \mathcal{L}_\phi + \mathcal{L}_\chi, 
    \label{eq:lagAll}
\end{equation}
where  $\mathcal{L}_{\text{SM}}$ represents the SM Lagrangian and
\begin{align}
   \mathcal{L}_{\zp} & =  g_{\zp} q_q  \sum_q  \bar{\psi}_q \gamma_\mu \psi_q \zpm -\frac{1}{4} F^{\prime \mu \nu} F_{\mu \nu}^{\prime}-\frac{1}{2} \sin{\epsilon}\, F^{\prime \mu \nu} B_{\mu \nu} \,, \label{eq:lagZprime}\\
   \mathcal{L}_{\phi} &= \left(\mathcal{D}^{\mu}\phi\right)^{\dagger}\left(\mathcal{D}_{\mu}\phi\right) - \mu_2^2 |\phi|^2 - \lambda_2 |\phi|^4 - \lambda_3 |\phi|^2 |H|^2 \,, \label{eq:lagS}\\
   \mathcal{L}_\chi &= \frac{i}{2} \overline{\chi} \cancel \partial \chi - \frac{1}{2} g_{\zp} q_\chi \zpm \overline{\chi} \gamma^5 \gamma_\mu \chi -\frac{1}{2} y_\chi \overline{\chi}\left(P_L \phi+P_R \phi^*\right) \chi \,. \label{eq:lagChi}
\end{align}
In the equations above, $P_{R,L} = \left(1\pm \gamma^5\right)/2$, while $\epsilon$ parametrizes the mixing angle between $\zp$ and the hypercharge gauge boson $B$.   
Since this parameter is highly constrained by experimental searches, we assume $\epsilon = 0$ in the following.
We also point out that the last term in Eq.~\eqref{eq:lagChi} is needed to give a mass to $\chi$ and requires $q_\phi = -2 q_\chi$.
Furthermore, it is interesting to note that the $\zp$-quark coupling is purely vectorial, while the $\zp$-DM coupling is purely axial.

After the dark scalar $\phi$ and the Higgs acquire vevs, three BSM mass eigenstates remain: $\zp$, $S$ and $\chi$. The scalar $S$ and the SM-like Higgs ($h$) correspond to linear combinations of the neutral components of the $\phi$ and $H$ fields:
\begin{align}
h &= H^0 \cos \alpha - \phi^0 \sin \alpha \,,  \nonumber\\
S &= \phi^0 \cos \alpha + H^0 \sin \alpha \,. \label{eq:alpha}
\end{align}
As a result, the couplings between $S$ and the SM fermions are proportional to $y_f \sin\alpha$, where $y_f$ is the corresponding SM Yukawa coupling. In particular, the $S{\rm-}t{\rm-}\bar{t}$ coupling ($y_t \sin\alpha$) is the dominant one and controls the $S$ production at the LHC.

The BSM masses are given by:
\begin{equation}
    m_{\zp} = 2 g_{\zp} q_{\chi} v_2\,,\; m_{S}^2 = m_h^2  + 2\frac{\lambda_{3}}{\sin2\alpha} v v_2\,, \mbox{ and } m_{\chi} = \frac{y_\chi}{\sqrt{2}} v_2\,,
\end{equation}
where $v/\sqrt{2}$ and $v_2/\sqrt{2}$ are the Higgs and $\phi$ vevs, respectively. The above relations allows us to write the $y_\chi$ coupling as a function of the $\zp$ and $\chi$ masses:
\begin{equation}
    y_{\chi} = 2 \sqrt{2}~ g_{\zp} q_\chi~ \frac{m_\chi}{m_{\zp}}\,. \label{eq:ychi}
\end{equation}
It is therefore convenient to take the following set of independent model parameters:
\begin{equation}
    \big\{m_{\zp},~ m_{S},~m_{\chi},~ g_{\chi},~ g_{q},~\sin{\alpha} \big\}, \label{eq:parameters}
\end{equation}
where $m_{\zp}$, $m_{S}$, $m_{\chi}$ are the $\zp$, $S$ and DM masses, respectively, $g_{\chi} \equiv g_{\zp} q_{\chi}$, $g_q \equiv g_{\zp} q_q$, 
and $\alpha$ is the $h$-$S$ mixing angle from Eq.~\eqref{eq:alpha}.

The phenomenology of the 2MDM model strongly depends on the masses of the BSM states. Since we want to investigate scenarios where $\zp$ and $S$ are allowed to decay to DM, but avoid constraints from invisible Higgs decays, we assume the following mass hierarchies:
\begin{equation}
    m_{h} < 2m_{\chi} < m_{\zp} ~ \mbox{  and } ~ m_{\chi} \leq m_{S} < m_{\zp}\,,
\end{equation}
where $m_{h}=125$~GeV is the observed Higgs mass.
More details are given in  Appendix~\ref{app:2mdm}.

\subsection{LHC signals} \label{sec:signal}

\begin{figure}[!t]
    \centering
    \includegraphics[width=0.95\textwidth]{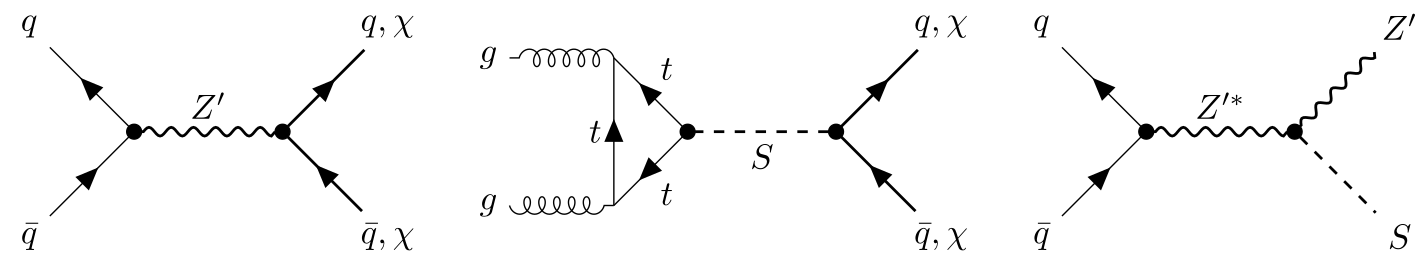}
    \caption{Leading order diagrams for the 2MDM signals at the LHC. The possible $\zp$ and $S$ decays are not indicated in the last diagram (associated production). \label{fig:diagrams}}
\end{figure}

The production of the new BSM states ($\zp,S,\chi$) at the LHC takes place mainly through the diagrams shown in Fig.~\ref{fig:diagrams}. While the associated $\zp S$ production illustrated by the third diagram can be relevant, it is always subdominant with respect to the on-shell (s-channel) production of $\zp$ and will be ignored in the following.
As we can see, the signal can be probed by searches for di-quark (dijet, $b\bar{b}$ and $t\bar{t}$) resonances and searches for jets plus $\etmiss$ with jets coming from initial state radiation (ISR).
In order to simulate the signal at the LHC, we have implemented the 2MDM model using FeynRules~\cite{Alloul:2013bka} and exported it to the UFO~\cite{Darme:2023jdn} format. 
The production cross-sections were calculated at leading order using \textsc{MadGraph5}~\cite{Alwall:2011uj} with the PDF set {\tt NNPDF31\_nnlo\_as\_0118} (see Ref.~\cite{smo3:dataset} for more details).

\begin{figure}[t!]  \centering    \includegraphics[width=0.6\textwidth]{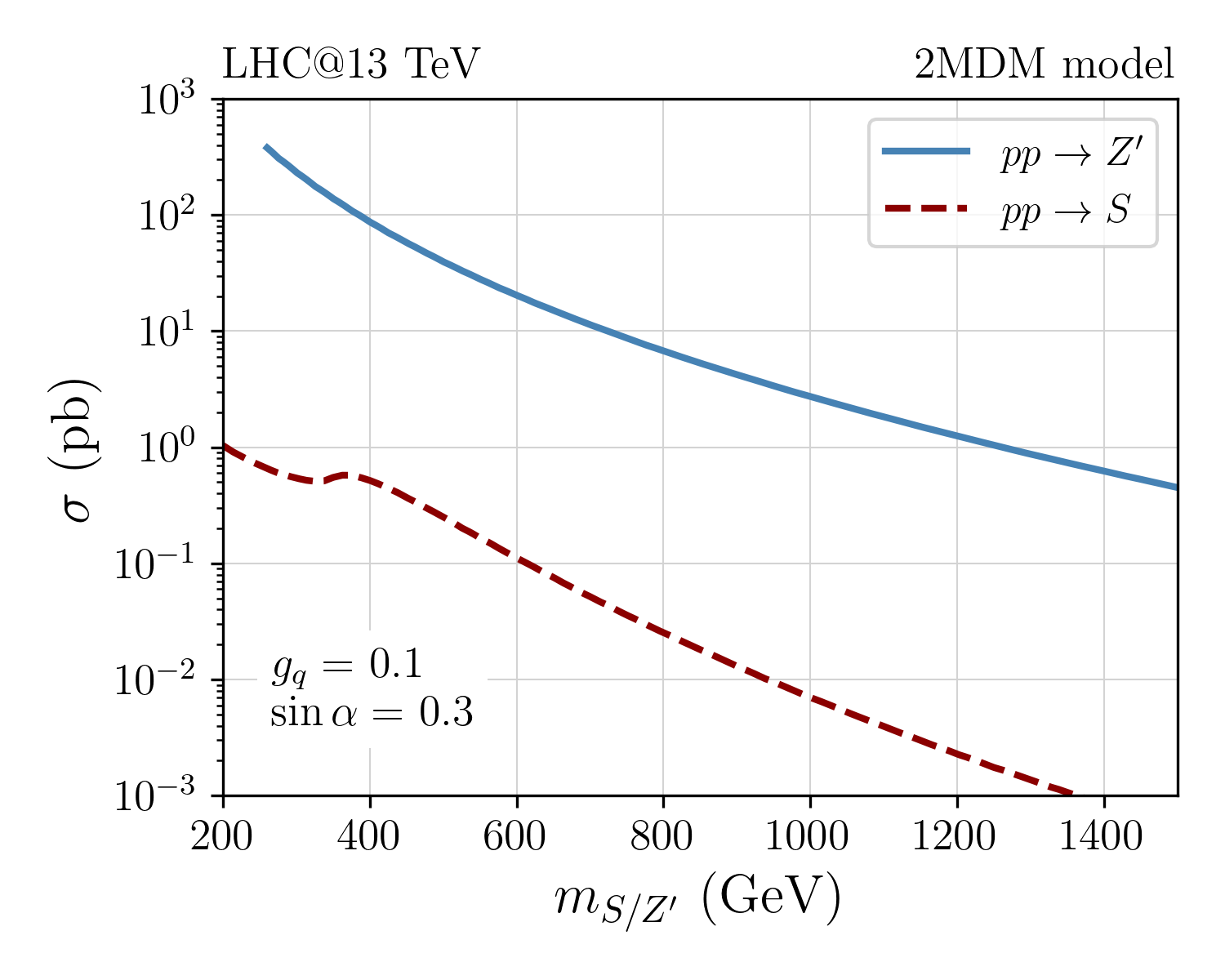}
    \caption{Cross-sections for the resonant production of the spin-1 ($\zp$, solid blue line) and spin-0 ($S$, dashed red line) mediators at the LHC. The $\zp$ coupling to quarks is fixed to $g_q = 0.1$, while the $h$-$S$ mixing angle is $\sin\alpha = 0.3$. Computed at leading order using \textsc{MadGraph5} \cite{Alwall:2011uj}. \label{fig:xsecs}}
\end{figure}

Table~\ref{tab:processes-couplings} shows how the cross-sections of the relevant processes scale with the model parameters. 
First, notice that the relative importance of the di-quark and $\etmiss$ channels depends on 
the ratio of $g_q/g_\chi$ for a $Z'$ 
mediator and $y_q/y_\chi$ for a $S$ mediator.
Second, the resonant production of $S$ is suppressed for small values of the mixing angle $\alpha$ and by a loop factor. As a result, the spin-0 production cross-section is typically much smaller than the spin-1 cross-section, unless $g_q \ll \sin\alpha$ and/or $m_{S} \ll m_{\zp}$.
This is illustrated in Fig.~\ref{fig:xsecs}.

\begin{table}[h!]
  \centering
  \caption{ \label{tab:processes-couplings} Dependence of the cross-sections of different processes from Fig.~\ref{fig:diagrams} (left and middle diagrams) on the model parameters from  Eq.~\eqref{eq:parameters} and $y_{\chi}$ defined in Eq.~\eqref{eq:ychi}.}
  \rowcolors{2}{gray!10}{white}
    \vspace{0.2cm}
  \begin{tabular}{l||c}
    \toprule
\hspace*{6mm}Process & Cross-Section \\
\toprule 
       $ p p \to  \zp \to q \overline{q}$  & $\sigma \propto g_{q}^4$ \\
       $p p \to \zp \to \chi \chi$ & $\sigma \propto g_{q}^2 g_{\chi}^2$\\
       $ p p \to  S \to q \overline{q} $ &  $\sigma \propto y_{t}^2 y_{q}^2 \sin^4{\alpha}$  \\
       $ p p \to S \to  \chi \chi$  & $\sigma \propto y_{t}^2 y_{\chi}^2 \sin^2{2 \alpha} $ \\      
       \bottomrule
    \end{tabular}
\end{table}

\begin{figure}[t!]    \centering
\includegraphics[width=0.65\textwidth]{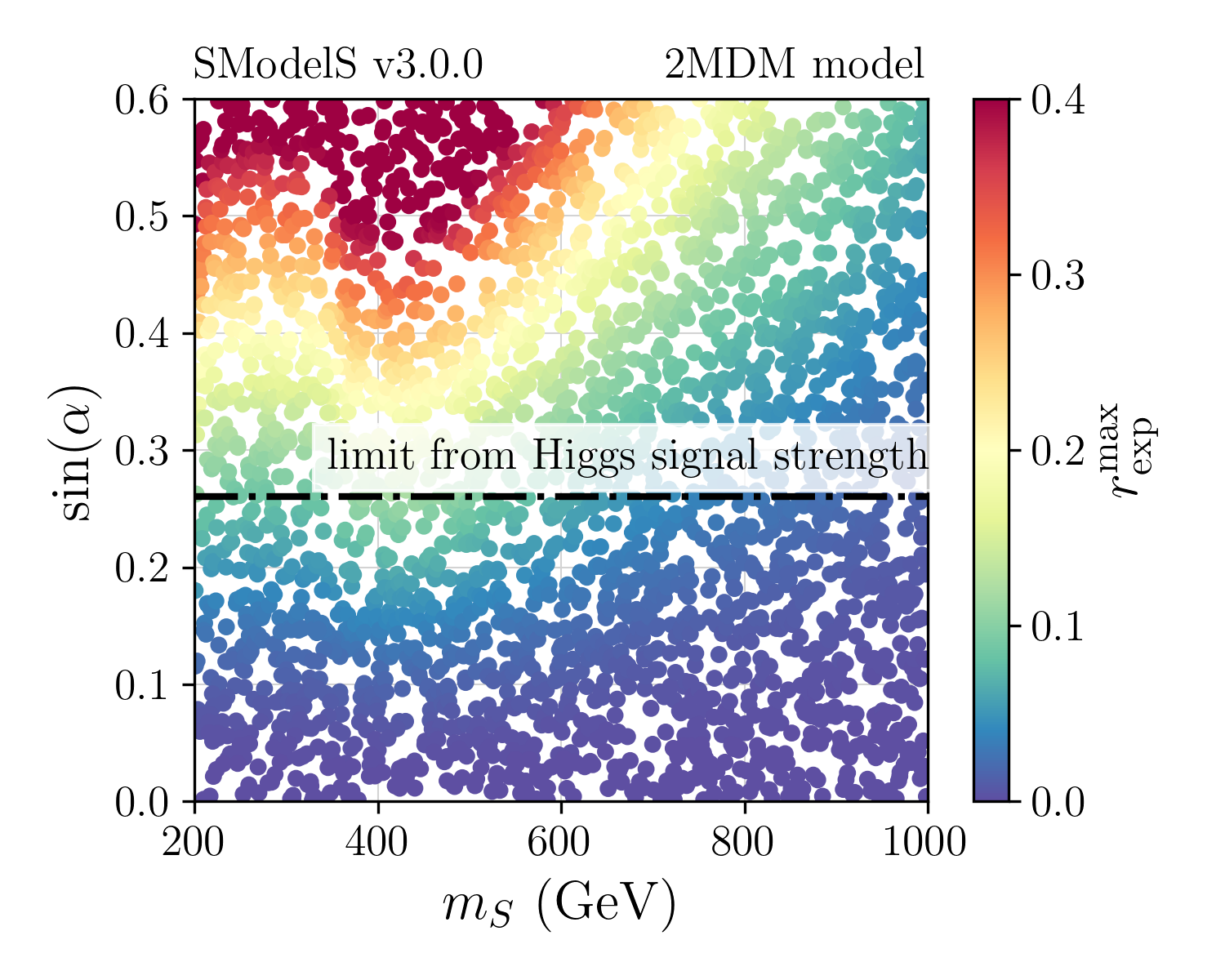}
    \caption{Ratio of $S$ production cross-section to its expected 95\% CL upper limit from CMS-EXO-20-004, $r_{\rm exp}^{\rm max} = \sigma(p p \to S)/\sigma_{\rm UL}^{\rm exp}$, in the $\sin\alpha$ vs.\ $m_S$ plane. 
    The dashed black line denotes the limit $\sin\alpha < 0.27$ from Higgs signal strength measurements.}
    \label{fig:2mdm_rs}
\end{figure}

The current limit on the $h$-$S$ mixing is $\sin\alpha < 0.27$~\cite{Ferber:2023iso} from SM Higgs signal strength measurements, assuming that
the SM-like Higgs does not decay into the dark sector. 
Even if we saturate this bound, the $S$ production cross-section (cf.\ Fig.~\ref{fig:xsecs}) is too small to be probed by resonance or $\etmiss$ searches. This lack of sensitivity is confirmed in Fig.~\ref{fig:2mdm_rs}, 
where we show the maximal attainable expected $r$-value, 
$r_{\rm exp}^{\rm max} = \sigma(p p \to S)/\sigma_{\rm UL}^{\rm exp}$, in the plane of $\sin\alpha$ vs.\ $m_S$.  
Here, $\sigma_{\rm UL}^{\rm exp}$ is the cross-section expected to be excluded at 95\% confidence level (CL) by the CMS monojet search (CMS-EXO-20-004).  
The above definition of $r_{\rm exp}^{\rm max}$ assumes BR$(S \to \chi \chi) = 100\%$ and thus quantifies the maximal sensitivity of the CMS search. 
We see from Fig.~\ref{fig:2mdm_rs} that the region of parameter space consistent with the Higgs measurements, $\sin\alpha < 0.27$, has $r_{\rm exp}^{\rm max} < 0.1$, i.e.\ much below the sensitivity of current monojet searches (an exclusion would require $r_{\rm exp}\ge 1$).
Furthermore, since decays to SM particles, e.g.\ $S \to t\bar{t}$, are further suppressed by $\sin^2\alpha$ (see Appendix~\ref{app:2mdm}), we do not expect any constraints from visible $S$ decays.

It is also worth mentioning that a recent search performed by ATLAS \cite{ATLAS-SUSY-2023-22} in the $t\bar{t}+\etmiss$ final state reports constraints for the associated production $p p \to S + t \bar{t}$ with $S \to \chi \chi$.
In this case, the cross-section is again suppressed for $\sin\alpha < 0.27$ and we have verified that the ATLAS analysis cannot exclude the 2MDM model for $m_S > 200$~GeV. 
Therefore, for the following results, we only include the constraints on $\zp$ production.

\subsection{Parameter scan and results} \label{sec:results}

To test the 2MDM model with \smothree, we performed a scan with 18k points
for distinct choices of $g_q$ and $g_{\chi}$ and covering
a wide range of $m_{\chi}$ and $m_{\zp}$ values. 
For definiteness, we have set $\sin\alpha = 0.25$ and $m_{S} = m_{\zp}/2$, though this has no impact on the results. 
The coupling values assumed as well as the ranges used for randomly scanning over the masses are shown in \cref{tab:2mdm-parameter}.
Note that the subset of parameters with $g_{\chi} = \sqrt{2}$ and $g_{q} = 0.25$ corresponds to the benchmark considered by ATLAS and CMS when presenting their limits within the simplified model framework.\footnote{The value of $g_\chi = \sqrt{2}$ in the 2MDM model corresponds to $g_\chi = 1$ in the simplified model scenario, since the latter assumes a Dirac DM particle. However, the simplified model benchmark assumes purely vectorial or purely axial $\zp$ couplings, while the 2MDM scenario has vectorial couplings to quarks and axial couplings to DM.}
For each set of parameter values, we compute the LO cross-sections for $\zp$ production at 8 and 13~TeV using \textsc{MadGraph5}~\cite{Alwall:2011uj} and the $\zp$ width and branching ratios using the expressions given in Appendix~\ref{app:2mdm}. 
These results are then converted to the \textsc{SLHA} format to be used as input for \smothree.
The average CPU time for running a single point in parameter space depends strongly on the analyses considered, and whether signal regions and/or analyses are combined. For the 2MDM model and the analyses listed in Table~\ref{tab:database}, the average running time per point using a single CPU are: $\sim 2$s without signal region combination, $\sim 26$s when combining signal regions and $\sim 5$min when combining signal regions and combining the CMS monojet (CMS-EXO-20-004) and the ATLAS multijet  (ATLAS-SUSY-2018-22) analyses.

\begin{table}[h]
    \centering
    \begin{tabular}{c|c|c|c|c}
        \hline
        $g_{q}$ & $g_{\chi}$ &  $m_{\chi}$ (GeV) & $m_{\zp}$ (GeV) & $N_{\rm points}$  \\ \hline 
        $0.1, 0.15$ & $0.01$ & 65 & (200, 3000) & 6k \\ 
        $0.15, 0.25$ & $1.0, \sqrt{2}$ & (65, $m_{\zp}/2$) & (200, 3000) & 6k \\
        $0.1$ & $0.6$ & (65, $m_{\zp}/2$)  & (200, 3000)   & 6k \\ \hline
         
    \end{tabular}
    \caption{Values and ranges for the relevant model parameters considered in this section. The $h$-$S$ mixing angle was taken as $\sin \alpha = 0.25$ and the scalar mass as $m_S = m_{\zp}/2$. The last column shows the number of scan points generated within each subset of parameters.}
    \label{tab:2mdm-parameter}
\end{table}

\begin{figure}[t]
    \centering
    \includegraphics[width=0.65\textwidth,clip]{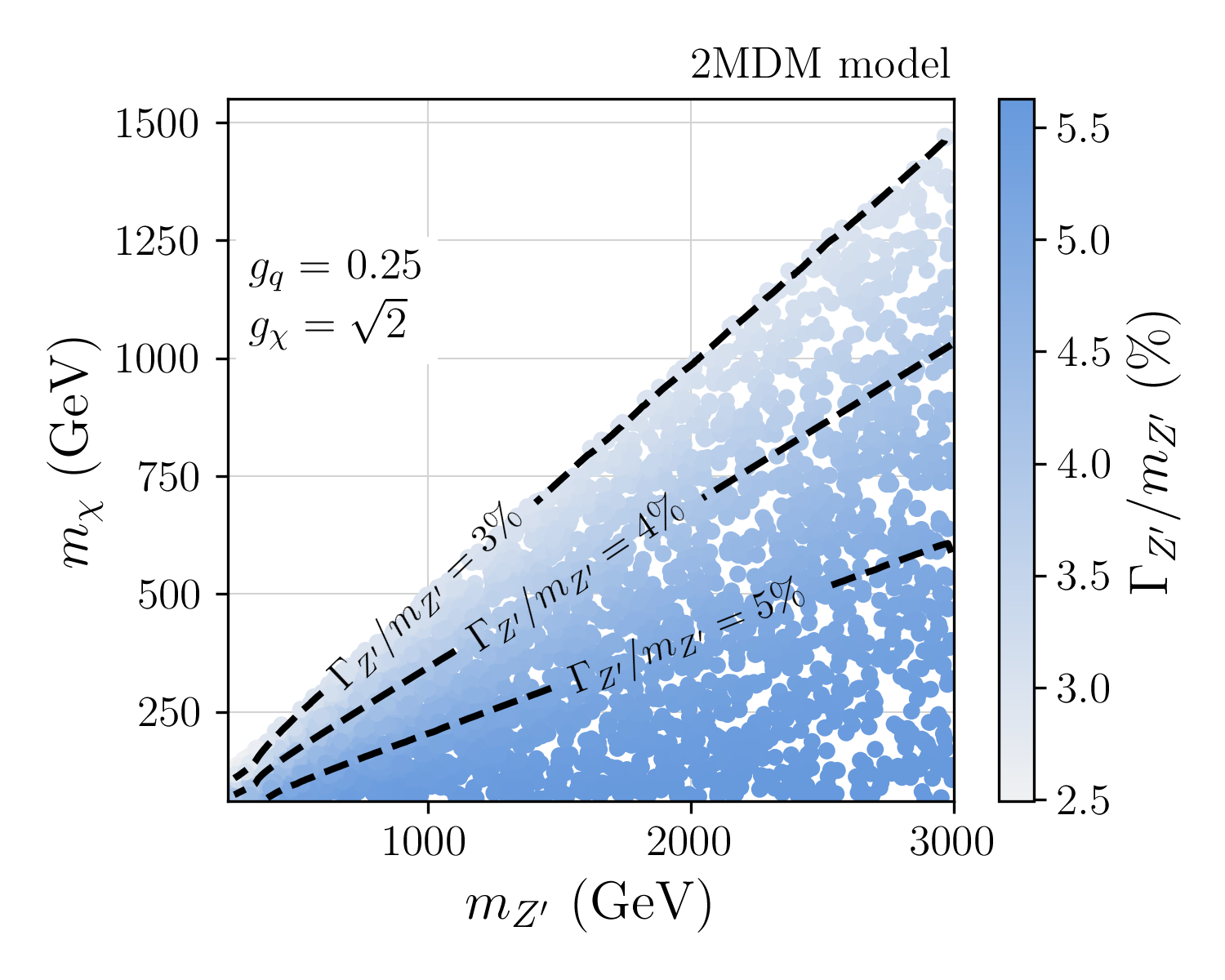}
    \caption{The $\zp$ width-to-mass ratio as a function of $m_{\zp}$ and $m_{\chi}$, for fixed values of the couplings: $g_q = 0.25$ and $g_\chi = \sqrt{2}$.}
    \label{fig:ratio-wm}
\end{figure}

Before turning to the \smodels results, a note on the $\zp$ width is in order. This width can be sizeable for sufficiently high values of $g_q$ and $g_\chi$, such that the NWA does not apply.  
As mentioned in the previous section, the only resonance search in the \smodels database which provides width-dependent results is CMS-EXO-19-012~\cite{CMS:2019gwf}. For the other resonance searches, the results can only be used if the NWA is applicable, i.e.\ if $\Gamma_{\zp}/m_{\zp} < 1\%$. 
The constraints from the jets$+\etmiss$ searches, on the other hand, are less dependent on the $\zp$  width~\cite{Chala:2015ama} 
and thus taken to be valid up to $\Gamma_{\zp}/m_{\zp} \simeq 5\%$. 
To illustrate the parameter space where this is relevant, Fig.~\ref{fig:ratio-wm} shows  $\Gamma_{\zp}/m_{\zp}$ as a function of $m_{\zp}$ and $m_\chi$, for $g_q = 0.25$ and $g_\chi = \sqrt{2}$ (i.e.\ for the maximal coupling values in our scan).
As can be seen, $\Gamma_{\zp}/m_{\zp}$ is always larger than 1\%, and, for regions where $m_{\chi} \leq m_{\zp}/5$, can reach up to $5.6\%$. 
Therefore, for this choice of couplings, the results of the $\etmiss$ searches are valid over the entire mass range considered; from the resonance searches, however, only the results from CMS-EXO-19-012 are applicable.

\subsubsection*{Constraints from jets plus $\etmiss$ searches}

We start the discussion of the results with the constraints obtained from the $\etmiss$ searches listed in Table~\ref{tab:database}.
These include the ATLAS and CMS monojet searches ATLAS-EXOT-2018-06~\cite{ATLAS:2021kxv} and CMS-EXO-20-004~\cite{CMS:2021far}, and the ATLAS multijet search ATLAS-SUSY-2018-22~\cite{ATLAS:2020syg}.
It is important to point out that, while the ATLAS multijet and the CMS monojet searches have provided the necessary information for computing a likelihood for the signal, the ATLAS monojet search has only provided upper limits and no statistical information.

\begin{figure}[t]\centering
    \includegraphics[width=0.66\textwidth]{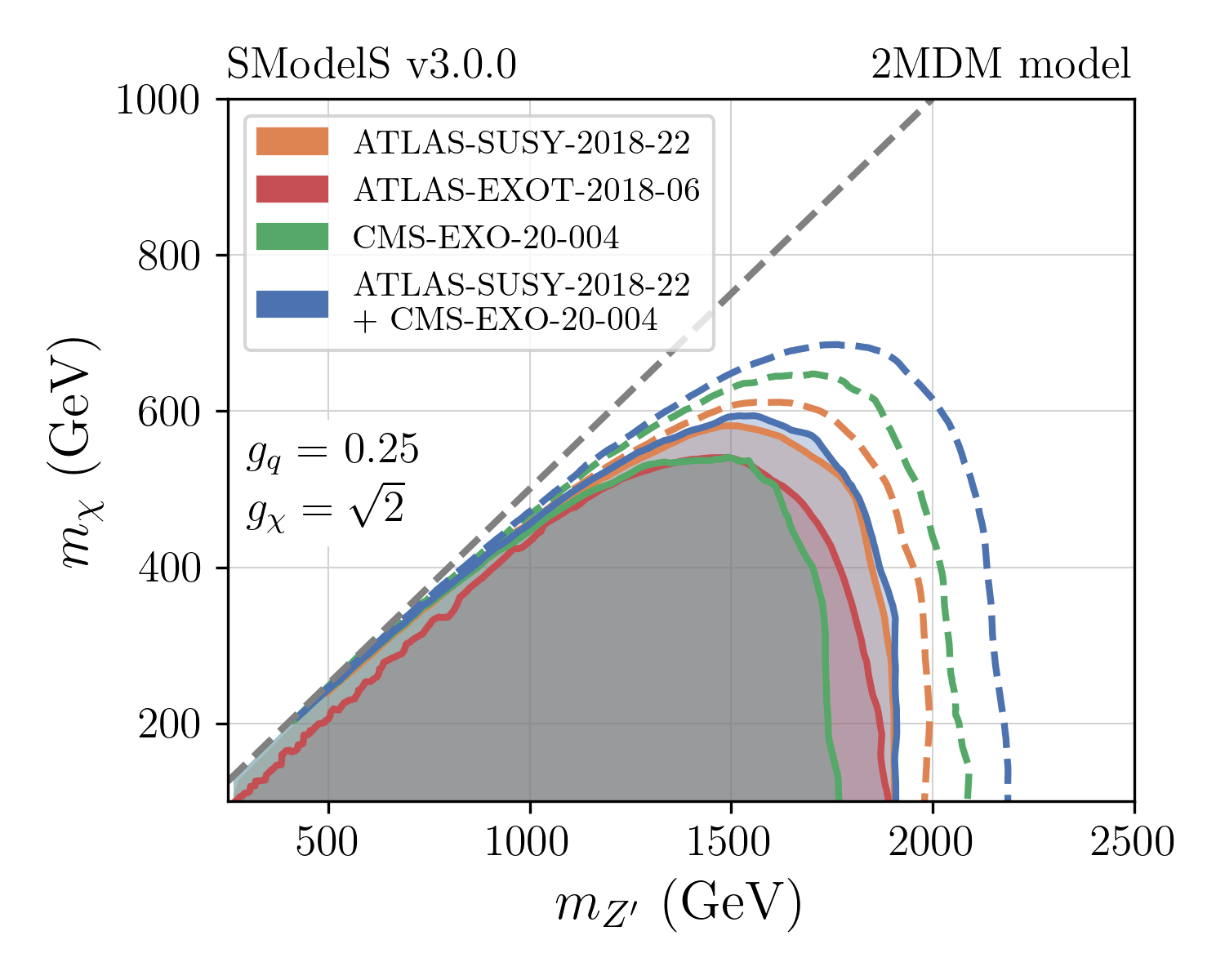}
    \caption{Exclusion curves for the monojet searches CMS-EXO-20-004 (green) and ATLAS-EXOT-2018-06 (red), the jets$+\etmiss$ search ATLAS-SUSY-2018-22 (orange), and the combination of CMS-EXO-20-004 and ATLAS-SUSY-2018-22 (blue). Observed exclusions are shown as solid lines, the expected ones as dashed lines. The dashed grey line indicates $2m_{\chi} = m_{\zp}$.}
    \label{fig:monojet-combined}
\end{figure}

Figure~\ref{fig:monojet-combined} shows the observed and expected exclusions obtained with \smothree in the plane of $m_\chi$ vs.\ $m_{\zp}$, for $g_q = 0.25$ and $g_\chi = \sqrt{2}$. 
As mentioned, these coupling values correspond to those adopted by ATLAS and CMS for their simplified model interpretations. 
Let us first consider the limits from the individual analyses (orange, red and green curves). The first thing to note is that for both the ATLAS multijet and the CMS monojet searches the observed limit is weaker than the expected one, meaning that the observed data exceeded the expected SM backgrounds.
While the ATLAS multijet search has observed up to $\sim 2.5\sigma$ excesses in two of its BDT signal regions, CMS has observed excesses of $\sim 3$--$4\sigma$ in several of its $\etmiss$ bins.\footnote{This is also true for the ATLAS monojet analysis, see Fig.~5a in \cite{ATLAS:2021kxv}; however, the analysis did not publish the expected $\sigma_{\rm UL}$ values, so we 
cannot determine the corresponding expected limit in Fig.~\ref{fig:monojet-combined}.}
The highest sensitivity, i.e.\ the strongest expected limit, 
comes from the CMS monojet search (shown in green). At the same time, this search provides the weakest observed limit, meaning it observed larger over-fluctuations than the other searches. 
In the region of small DM masses, mediator masses up to 1.9~TeV are excluded by the ATLAS analyses (orange and red curves), while the CMS monojet search excludes up to 1.7~TeV (green curve).
It is also worth noting that for higher DM masses, the ATLAS monojet search (red curve) is slightly less constraining than the multijet search (orange curve).
Finally, note that for our choice of coupling values, we have BR$(\zp \to \chi \chi) \simeq 50\%$ for $m_\chi \ll m_{\zp}$. This BR decreases with increasing $m_\chi$, resulting in the loss of sensitivity close to the $2m_\chi = m_{\zp}$ line (grey dashes).

For EM-type results, \smodels can compute not only $r$-values (to see whether a point is allowed or excluded) but also likelihoods. This allows for much more sophisticated 
statistical interpretation; for instance, one may derive the exclusion CLs, perform hypothesis tests and/or compute $p$-values. 
Moreover, the likelihoods of statistically independent analyses can be combined, leading to more robust constraints~\cite{MahdiAltakach:2023bdn}. 
This brings us to the blue curves in Fig.~\ref{fig:monojet-combined}, which show the mass limits obtained from the combination of the ATLAS-EXOT-2018-22 and CMS-EXO-20-004 analyses.  
Comparing the
dashed lines, we see that the combination 
extends the {\it expected} reach by 100--200~GeV beyond that of the individual analyses (from $m_{\zp} \approx 2$~TeV for ATLAS and 2.1~TeV for CMS to about 2.2~TeV for the combination). The combined {\it observed} exclusion (solid blue curve), however, while stronger than the CMS monojet limit, is almost the same as the ATLAS multijet limit alone:  
the effect of the combination is that tensions between the ATLAS and CMS data are levelled out and one obtains a statistically more robust limit.

\begin{figure}[t] \centering
\includegraphics[width=1.0\linewidth]{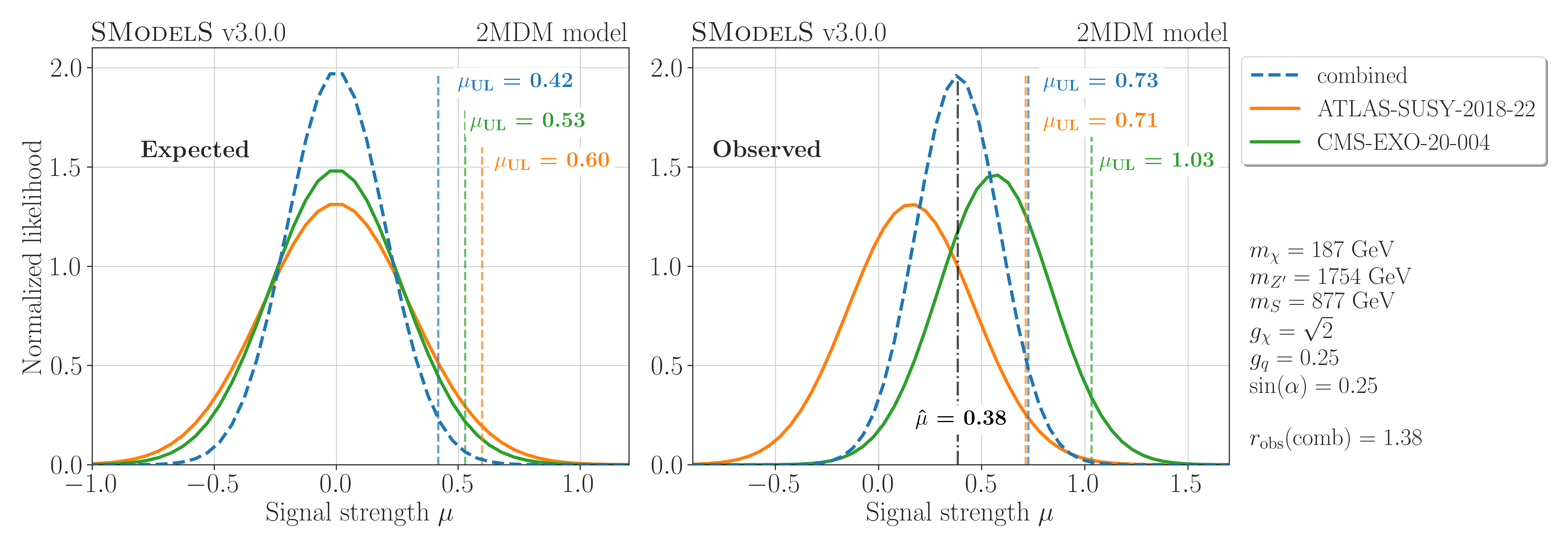}
    \caption{Expected (left) and observed (right) normalized likelihoods for the ATLAS-SUSY-2018-22 (multijet) and CMS-EXO-20-004 (monojet) analyses and their combination, for a sample point of the 2MDM model with $m_{\chi} = 187$~GeV and $m_{\zp} = 1754$~GeV, as a function of the signal strength $\mu$.}
    \label{fig:likelihoods}
\end{figure}

In order to better understand this behaviour, we plot in Fig.~\ref{fig:likelihoods} the expected and observed likelihoods as a function of the signal strength ($\mu$) for the individual searches as well as for their combination. The masses are fixed to $m_{\chi} = 187$ GeV and $m_{\zp} = 1754$ GeV, which corresponds to a point excluded by the ATLAS multijet search, but not by the CMS monojet search. 
The left plot in Fig.~\ref{fig:likelihoods} (expected likelihood) shows how the sensitivity improves with the combination of analyses. Concretely, since the $r$-values are inversely proportional to the upper limits on the signal strength, $r=1/\mu_{\rm UL}$, we go from 
$r_{\rm exp}=1.67$ and 1.87 for the individual analyses to $r_{\rm exp}=2.38$ for the combination.
The right plot shows the observed likelihoods.
As already discussed, both the ATLAS and the CMS analyses favour $\mu > 0$, due to small excesses in the data.
The observed over-fluctuations being larger in CMS than in ATLAS,
the CMS likelihood is maximized for $\mu \simeq 0.58$, while the ATLAS one is maximal at $\mu \simeq 0.18$. This results in a combined likelihood (dashed curve) which peaks at $\mu \simeq 0.38$. The corresponding 95\%~CL limits are $\mu_{\rm UL}=1.03$, $0.71$ and $0.73$ for the CMS and ATLAS analyses and their combination, respectively. 
We see that indeed ATLAS excludes the point, but CMS does not. The combination of the two analyses centres the likelihood in-between the ATLAS and the CMS ones, and narrows it to a smaller standard deviation. The result is a solid exclusion with $r_{\rm obs}=1.38$, which is in fact slightly weaker than for ATLAS alone.

\begin{figure}[t]
    \centering
    \includegraphics[width=0.6\textwidth]{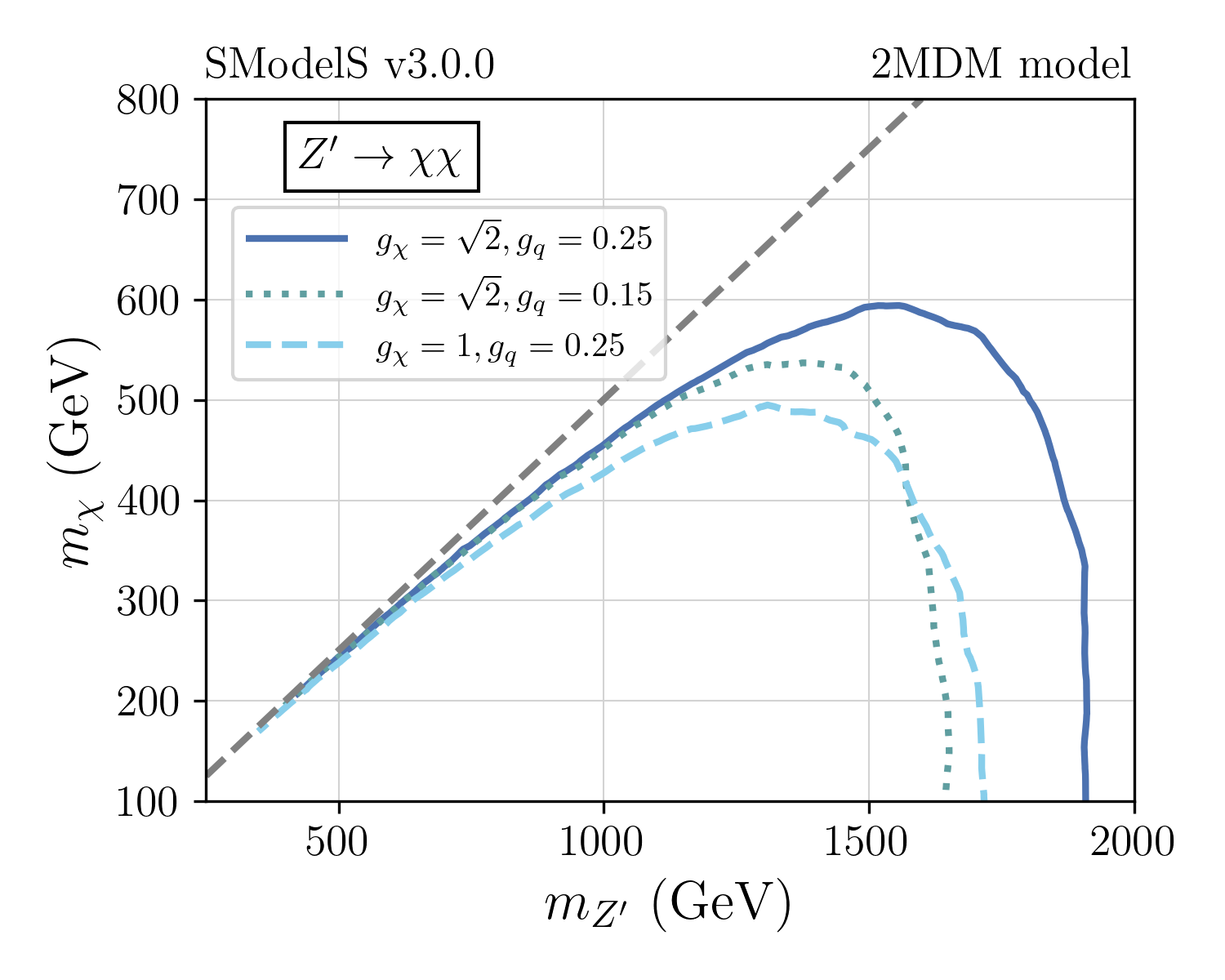}
    \caption{Exclusion lines in the $m_{\chi}$ vs.\ $m_{\zp}$ plane from the combination of the ATLAS multijet and the CMS monojet searches, for three different choices of $g_{q}$ and $g_{\chi}$.}
    \label{fig:2mdm_monojet}
\end{figure}

Next, we consider how the results from Fig.~\ref{fig:monojet-combined} depend on $g_q$ and $g_\chi$.
In the NWA and for $m_{\chi} \ll m_{\zp}$, the signal in the $\etmiss$ channel is given by
\begin{equation}
    \sigma(p p \to \zp)\times {\rm BR}(\zp \to \chi \chi) =  \frac{\sigma(p p \to \zp) \, \Gamma(\zp \to \chi \chi)}{\Gamma(\zp \to \chi \chi) + \Gamma(\zp \to q \bar{q})} \, \propto\, g_q^2 \frac{1}{1+g_q^2/g_\chi^2}\,. \label{eq:metsignal}
\end{equation}
We see from this expression that, for fixed $g_q$, the \etmiss signal increases with $g_{\chi}$. Likewise, for fixed $g_{\chi}$, the signal increases with $g_q$, because the increase in the production cross-section makes up for the decrease in the invisible branching ratio.
Altogether, the signal increases or decreases both with $g_q$ and $g_\chi$.
This is illustrated in Fig.~\ref{fig:2mdm_monojet}, which shows the combined exclusion from the ATLAS multijet and CMS monojet searches for distinct values of $g_q$ and $g_\chi$. 
The dashed and dotted blue curves show that reducing either $g_q$ or $g_{\chi}$ with respect to the values used in Fig.~\ref{fig:monojet-combined} results in a weaker exclusion.

\subsubsection*{Di-quark resonance searches}

In addition to the jets plus $E_T^{\rm miss}$ signal, the 2MDM model can also be probed by resonant searches, as BR$(\zp \to q\bar{q})$ can be significant for $g_q \gtrsim g_{\chi}$ and/or $m_{\zp} \sim m_{\chi}/2$, since along the kinematical edge ($m_{\zp} = m_{\chi}/2$) the invisible decay is suppressed. 
However, as discussed above, most of the resonant searches only provide limits under the NWA, which can be violated for sufficiently large $g_q$ and/or $g_{\chi}$ (see Appendix~\ref{app:2mdm}).

Figure~\ref{fig:exclusions_2mdm_dijet} shows the $r_{\rm obs}$ values obtained for the various resonance searches listed in \cref{tab:database} as a function of $m_{\zp}$. 
The plot on the left shows the limits from dijet resonance searches ($\zp \to j j$), while the one on the right shows those from $\zp \to b\bar{b},t\bar{t}$.\footnote{In the left plot, the ATLAS-EXOT-2019-03 and CMS-EXO-19-012 constraints also include the (subdominant) contribution from $\zp \to b\bar{b}$, since these searches also allow for decays to $b$ quarks.}
Here, we have set $g_q = 0.1$, $g_\chi = 0.01$ and $m_\chi = 65$~GeV, so the  $\zp$ decays almost exclusively to light jets, $b\bar b$ and top pairs (if kinematically allowed).  Moreover, $\Gamma_{\zp}/m_{\zp}\lesssim 0.5\%$ for the whole mass range considered, thus satisfying the NWA. 
The low mass region ($m_{\zp} < 1.5$~TeV) is covered by the 8 TeV searches, with CMS-EXO-16-057 ($b\bar b$ final state) excluding $\zp$ masses around 300~GeV and 500~GeV and ATLAS-EXOT-2013-11 (dijet final state) excluding the range $0.8 \mbox{ TeV} < m_{\zp} < 1.35$~TeV, except for a small mass window between 1~TeV and 1.2~TeV, where over-fluctuations were observed, reducing the constraining power in this region.
The coverage of the 13~TeV searches starts at 1.5~TeV and represents a significant increase in sensitivity, as shown by the ATLAS-EXOT-2019-03 (dijet, red curve) limit. As a result, the exclusion power greatly increases once the 13~TeV searches kick in and $1.5 \mbox{ TeV} < m_{\zp} < 2.4$~TeV is excluded.
For larger $g_q$ values, the $\zp$ production cross-section increases, but $\Gamma_{\zp}$ also grows, resulting in a broad resonance for large $m_{\zp}$. 
Since only the CMS-EXO-19-012 dijet resonance search provided limits for large widths, this is the only analysis applicable beyond the NWA. The dotted curves in Fig.~\ref{fig:exclusions_2mdm_dijet} illustrate the exclusion obtained for $g_q = 0.15$ with the other parameters unchanged.
In this case, $\Gamma_{\zp}/m_{\zp}>1\%$ for $m_{\zp} > 300$~GeV,
and it is no longer safe to assume the NWA.
As a result, only the region $m_{\zp} < 300$~GeV is excluded by CMS-EXO-16-057 ($b\bar b$), see the dotted magenta curve.
In the high mass region, the dijet resonance search CMS-EXO-19-012
excludes $1.8 \mbox{ TeV} < m_{\zp} < 2.5 \mbox{ TeV}$. We expect that lower masses can also be probed by this and other searches, but unfortunately, the necessary information is not publicly available to reliably constrain $\zp$ masses between 300~GeV and 1.8~TeV for large widths.

\begin{figure}[t!] 
    \includegraphics[width=0.52\textwidth]{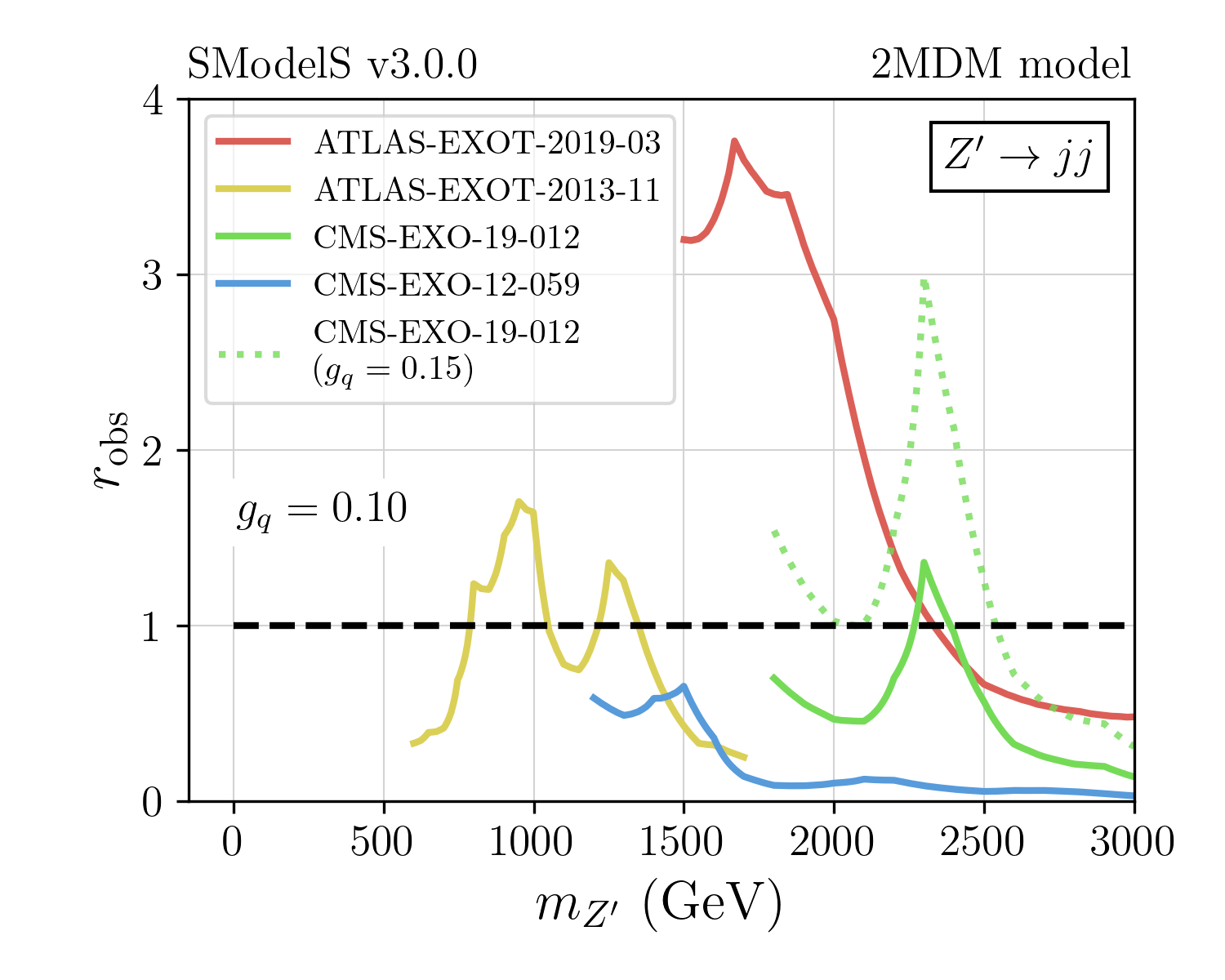}\hspace*{-4mm}\includegraphics[width=0.52\textwidth]{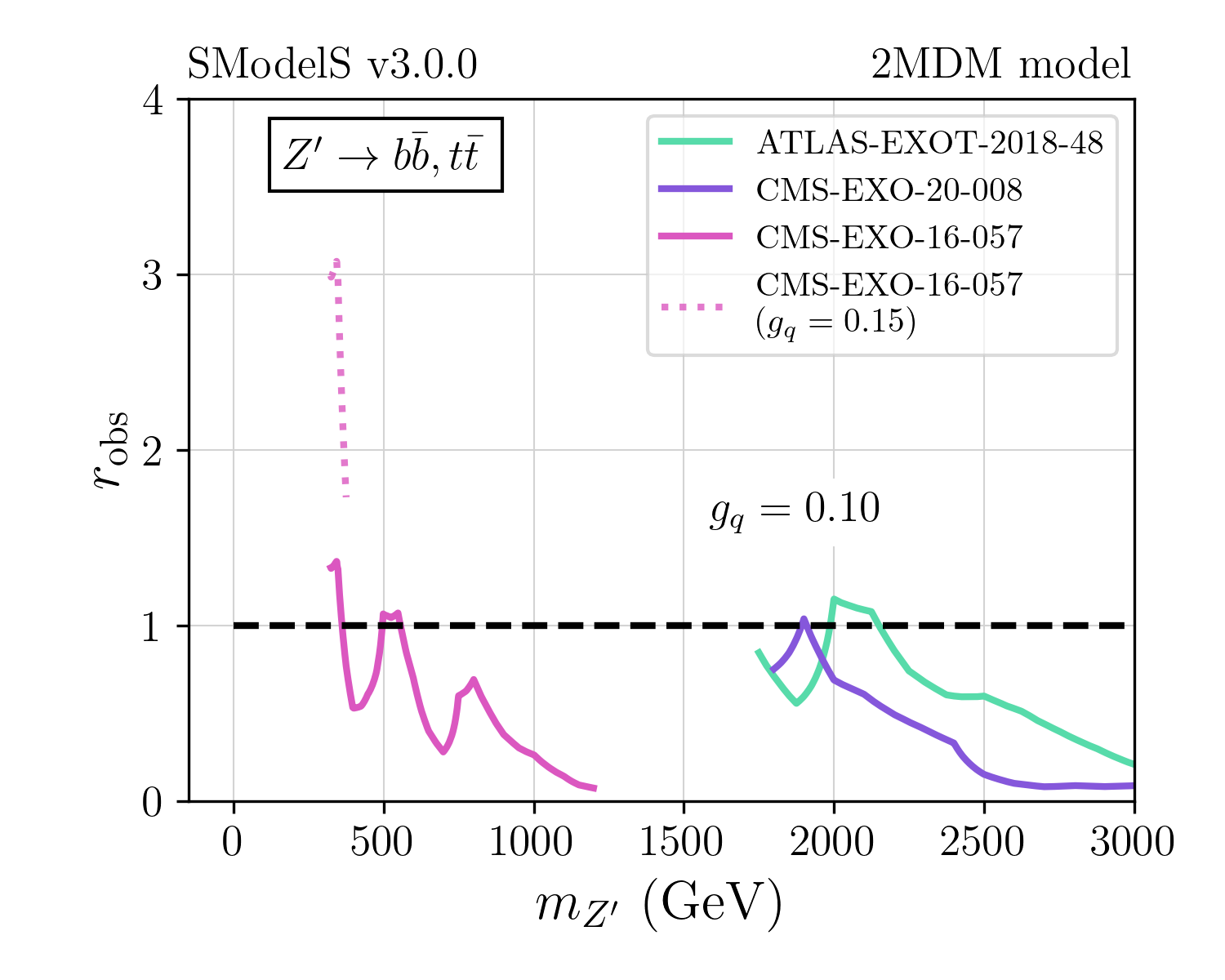}
    \caption{Observed $r$-value versus $m_{\zp}$ for various ATLAS and CMS resonance searches, denoted in colour. The solid (dotted) lines correspond to $g_q = 0.1\ (0.15)$, with the $\zp$-DM coupling set to $g_{\chi} = 0.01$. For these values, the $\zp$ predominantly decays to quarks. The region above the dashed black line ($r_{\rm obs} > 1$) is excluded by the corresponding analysis.} 
    \label{fig:exclusions_2mdm_dijet}
\end{figure}

\begin{figure}[t!] 
     \hspace*{-2mm}\includegraphics[width=0.52\textwidth]{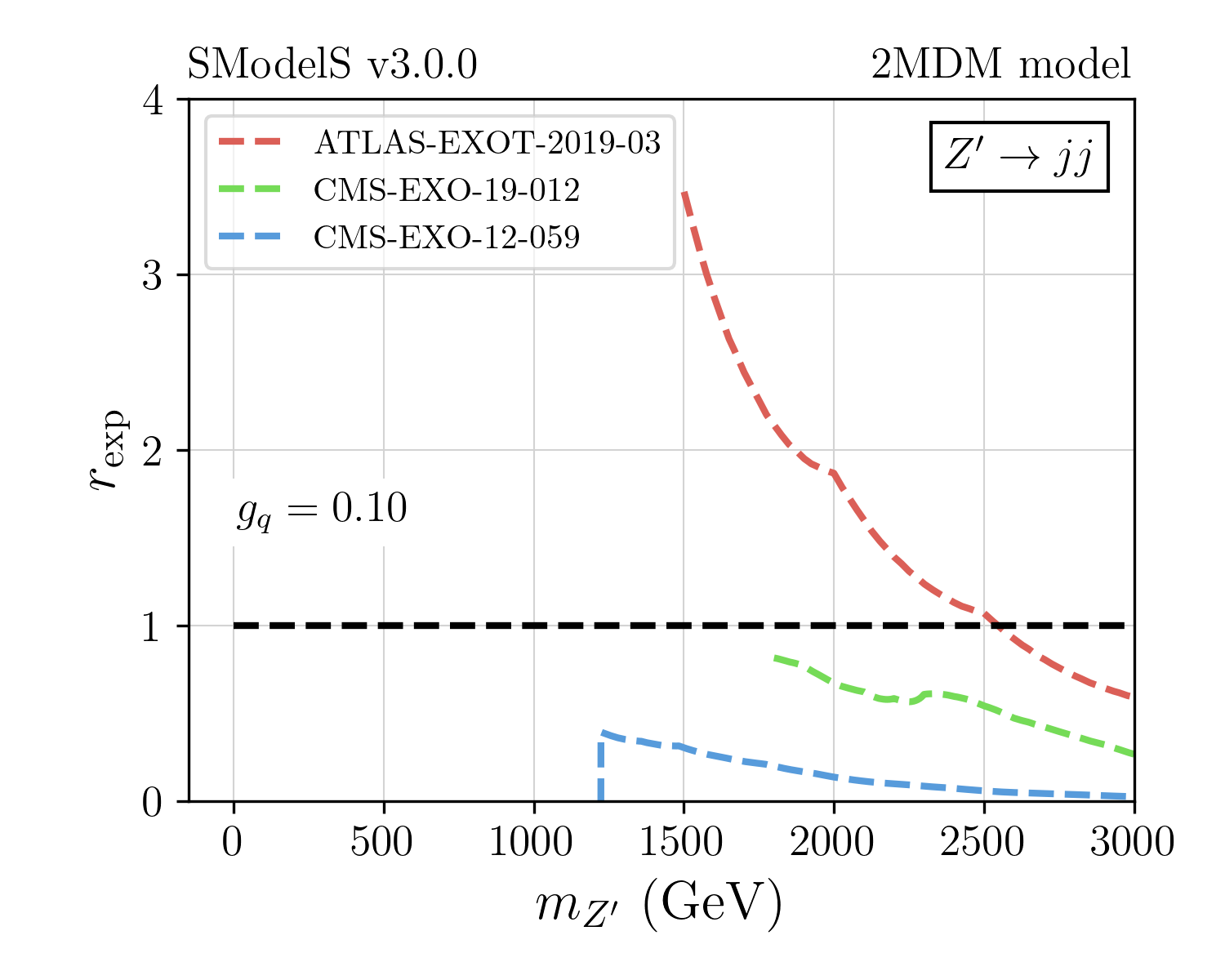}\hspace*{-4mm}\includegraphics[width=0.52\textwidth]{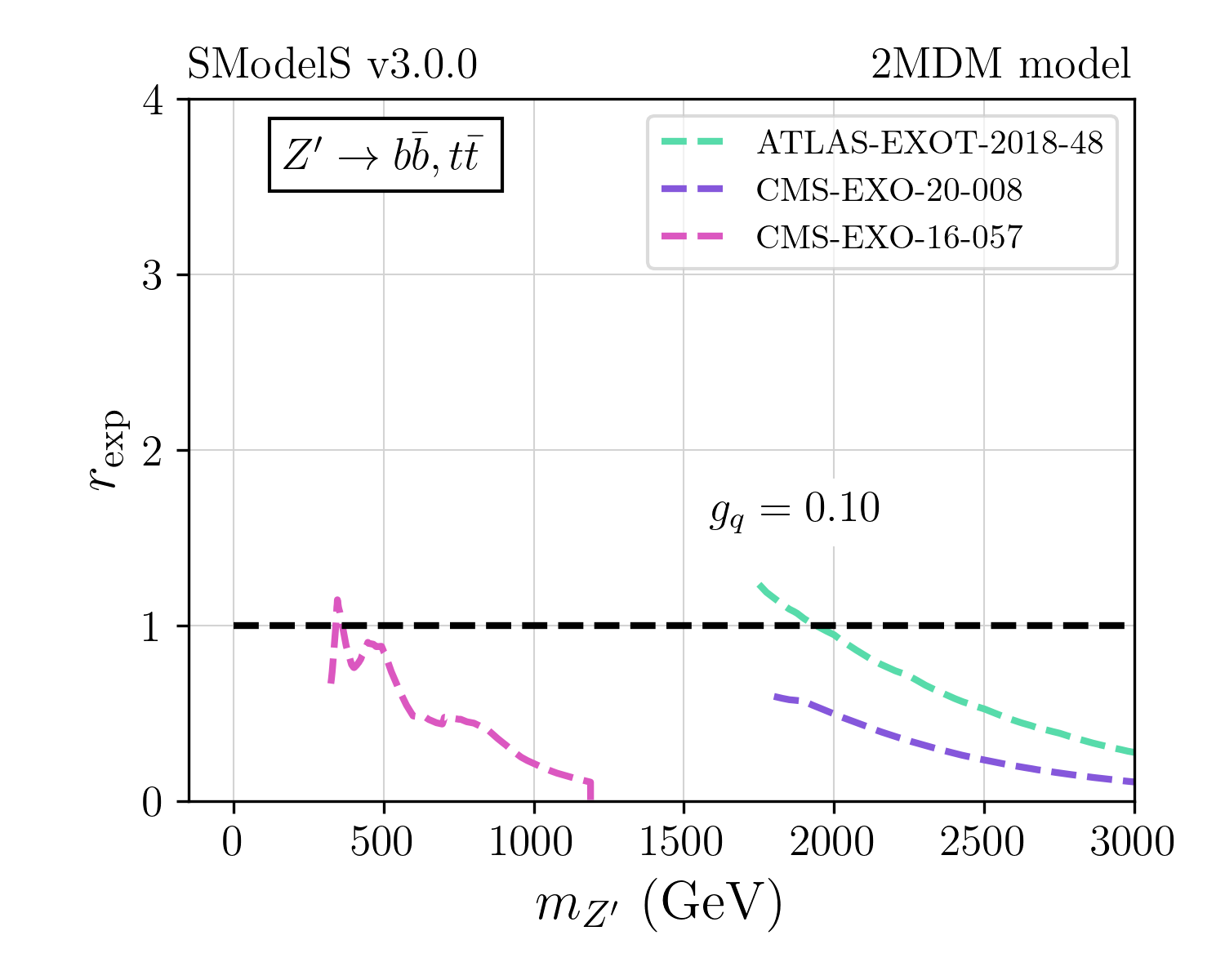}
    \caption{As Fig.~\ref{fig:exclusions_2mdm_dijet} but for the expected  $r$-values, 
    except for ATLAS-EXOT-2013-11 for which expected limits are not publicly available.} 
    \label{fig:exclusions_2mdm_dijet_expected}
\end{figure}

A word of caution is in order regarding up- and downward fluctuations in the resonance searches. These can be estimated from comparing the expected $r$-values, shown in Fig.~\ref{fig:exclusions_2mdm_dijet_expected}, to the observed ones in Fig.~\ref{fig:exclusions_2mdm_dijet}.
One conclusion from Fig.~\ref{fig:exclusions_2mdm_dijet_expected} is that ATLAS-EXOT-2019-03 is more sensitive than CMS-EXO-19-012, and ATLAS-EXOT-2018-48 is more sensitive than CMS-EXO-20-008; the overall most sensitive analysis in the high-mass range is ATLAS-EXOT-2019-03. Another conclusion is that fluctuations are ubiquitous; it would be highly interesting to be able to combine at least the ATLAS and CMS dijet resonance searches to average out 1--2$\sigma$ effects and obtain more robust limits in the region of $m_{\zp}\lesssim 2.5$~TeV.

\subsubsection*{Resonance vs $\etmiss$ searches}

\begin{figure}[t]
    \hspace*{-2mm}\includegraphics[width=0.52\textwidth]{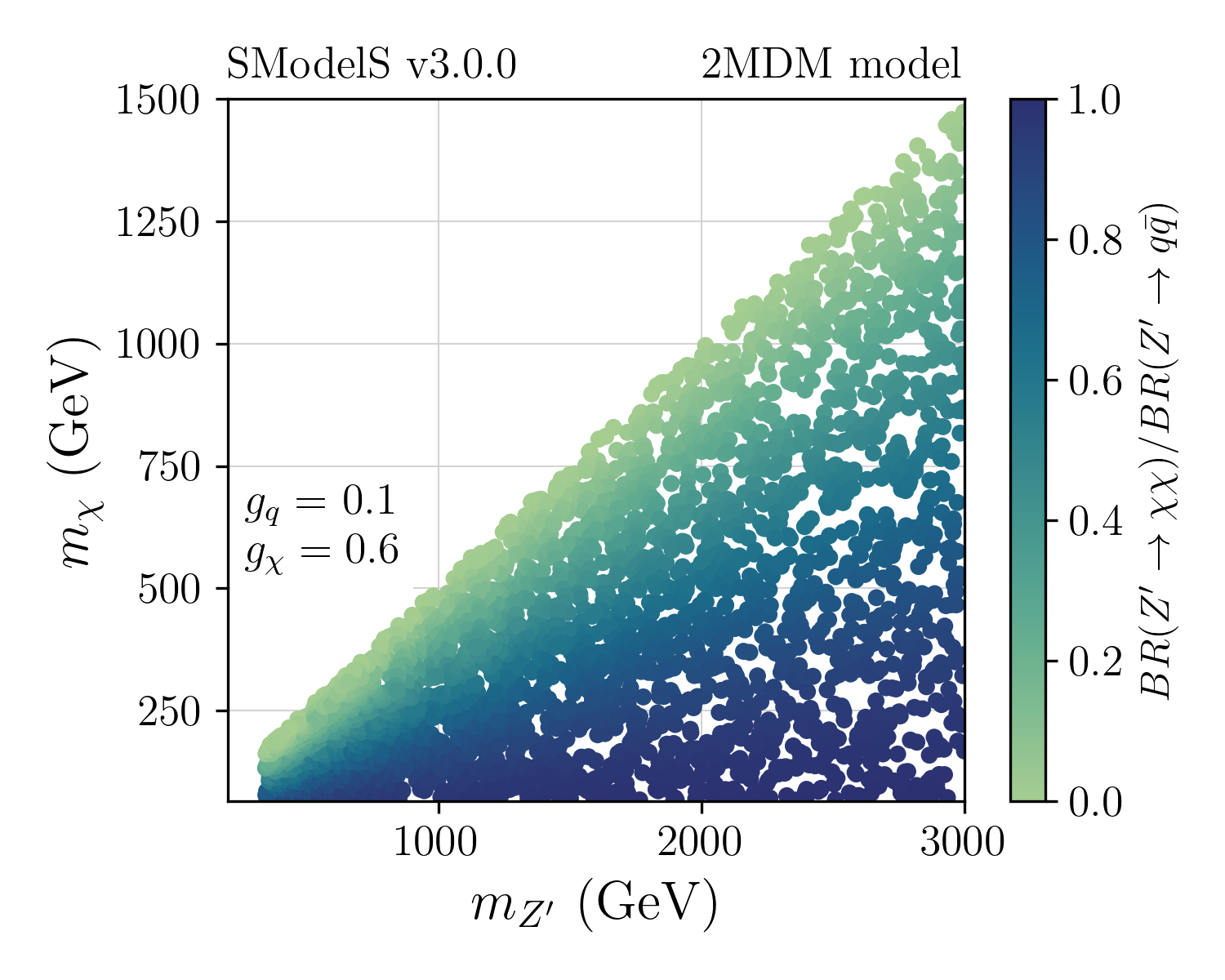}%
    \includegraphics[width=0.52\textwidth]{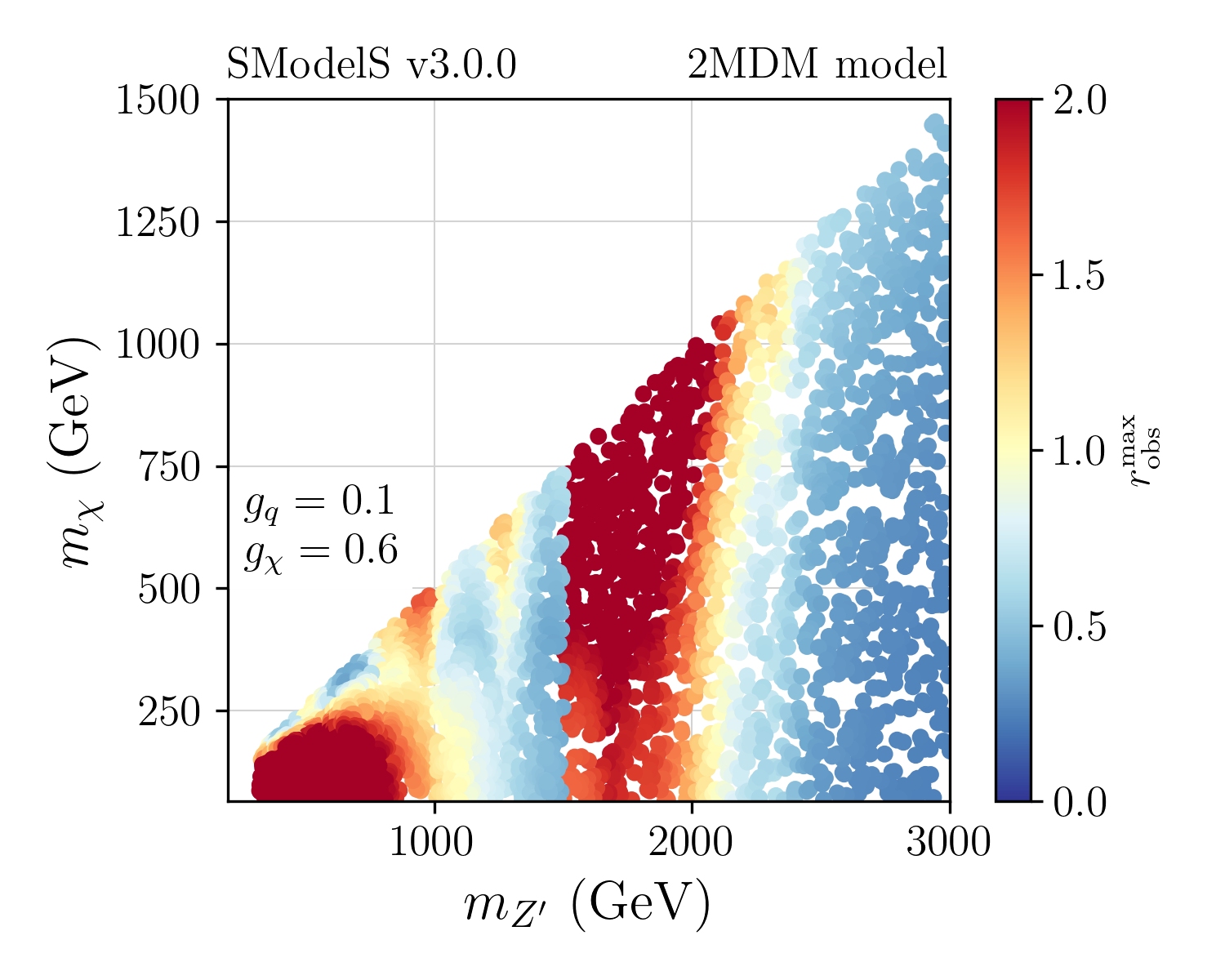}\\
    \hspace*{-2mm}\includegraphics[width=0.52\textwidth]{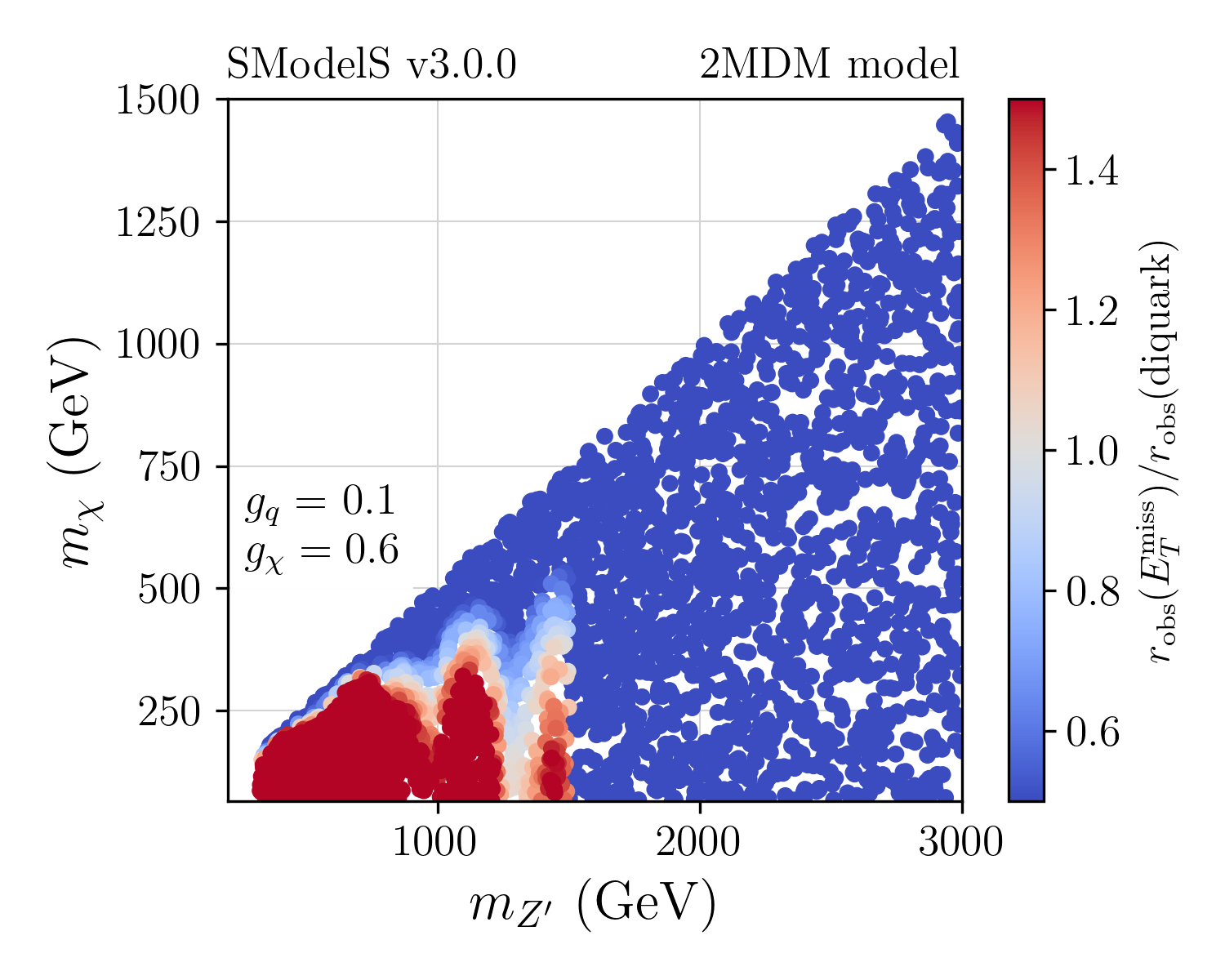}%
    \includegraphics[width=0.52\textwidth]{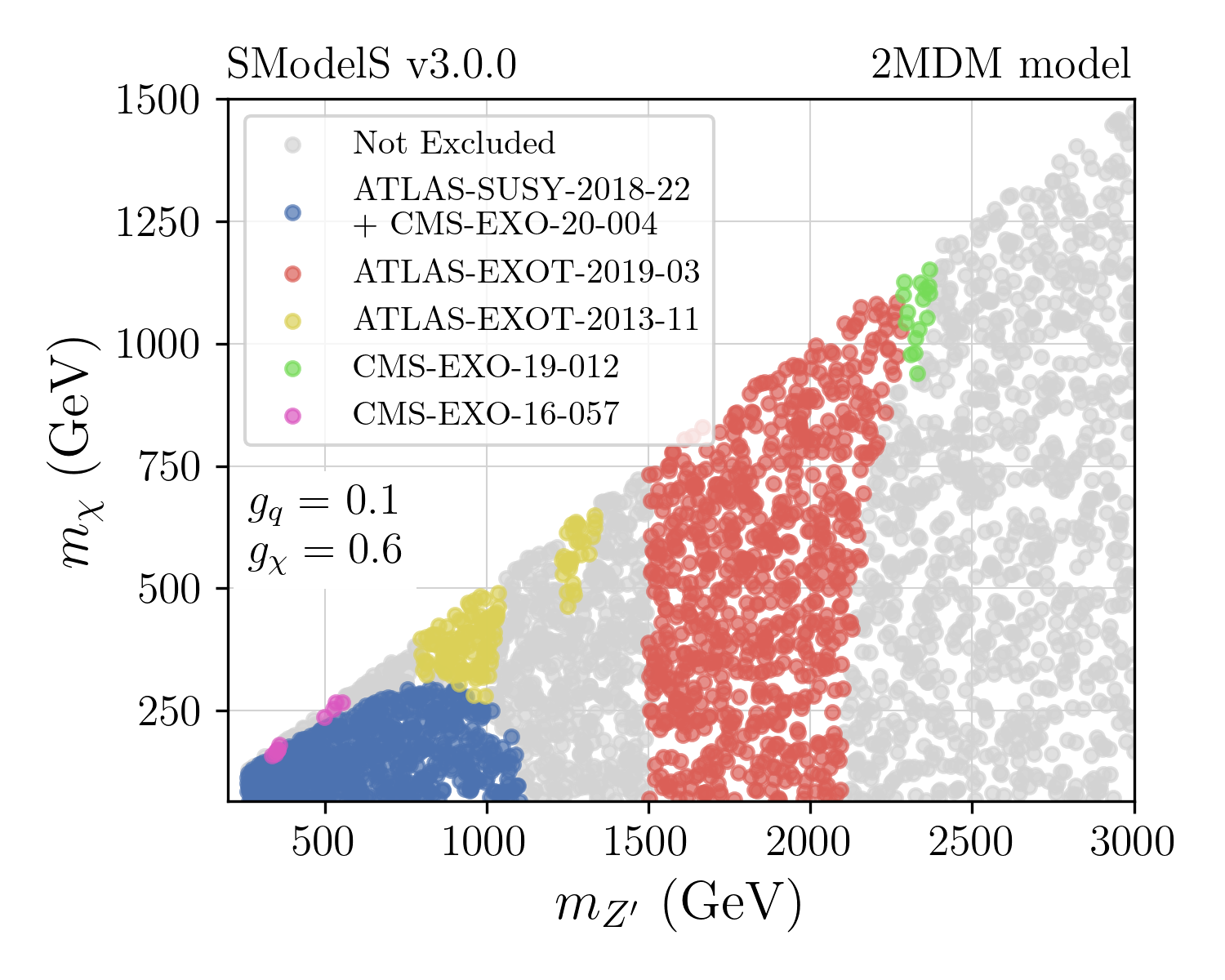}
    \caption{Results in the $m_\chi$ vs.\ $m_{\zp}$ plane for $g_q=0.1$ and $g_\chi=0.6$. {\it Top left}: Ratio of branching ratios of $\zp\to\chi\chi$ and $\zp\to q\bar q$ decays. {\it Top right}: maximal observed $r$-value from the combination of \etmiss\ searches (ATLAS-SUSY-2018-22 plus CMS-EXO-20-004) or individual resonance searches.
    {\it Bottom left}: Ratio of $r_{\rm obs}$ from the combined $\etmiss$ searches, labelled $r_{\rm obs}(\etmiss)$, to the largest $r_{\rm obs}$ from resonance searches, labelled $r_{\rm obs}$(diquark).
    {\it Bottom right}: allowed (grey) and excluded (coloured) points. For each excluded point the colour corresponds to the analysis with the largest $r_{\rm obs}$. }
    \label{fig:ratio-dijet-monojet}
\end{figure}

It is also interesting to compare the relative impact of resonance and $\etmiss$ searches 
on the 2MDM scenario, as done in Fig.~\ref{fig:ratio-dijet-monojet} for the case $g_q = 0.1$ and $g_\chi = 0.6$.
\footnote{The complementarity between resonance and \etmiss searches has been explored previously in the literature, see e.g.~\cite{Chala:2015ama,Arina:2016cqj,Kraml:2017atm,Albert:2017onk,Albert:2022xla,ATLAS:2024kpy}.}
The sensitivity of each search depends on the visible and invisible branching ratios, BR$(\zp \to \chi \chi)$ vs.\ BR$(\zp \to q \bar{q})$, and on the $\zp$ width, since for broad resonances the applicability of di-quark searches is limited. The ratio of invisible over visible decay BRs as a function of $m_\chi$ vs.\ $m_{\zp}$ is shown explicitly in the top left panel in Fig.~\ref{fig:ratio-dijet-monojet}; the choice of couplings made here results in a small $\zp$ width, hence the NWA is valid all across the plane. 

The top right and bottom left panels show, respectively, the highest observed $r$-value  ($r_{\rm obs}^{\rm max}$) and   
the ratio of $r_{\rm obs}$ values from $\etmiss$ searches to those from resonance searches. 
For the $\etmiss$ analyses, we consider the combination of the ATLAS multijet and CMS monojet searches, since this gives the most sensitive constraints.
For the resonance (diquark) searches, on the other hand, lacking expected ULs, we use the analysis with the largest $r_{\rm obs}$ for each point in parameter space. 
The $r_{\rm obs}^{\rm max}$ plot displays the level to which different phase space regions are challenged by the experimental results (recall that $r_{\rm obs}^{\rm max}=1$ corresponds to the observed 95\%~CL exclusion).
The $r_{\rm obs}$ ratio, on the other hand, illustrates the relative exclusion power of each type of search as a function of the DM and mediator masses. 
As we can see, for $m_{\zp} < 1.2$~TeV the $\etmiss$ searches are dominant (red points), except for $m_{\chi} \sim m_{\zp}/2$, where the invisible $\zp$ decay becomes kinematically suppressed.
For $m_{\zp} > 1.5$~TeV the resonance searches rapidly take over
(blue points), since for these masses there are limits from the
dijet search ATLAS-EXOT-2019-03, which has a high constraining power, as shown in Fig.~\ref{fig:exclusions_2mdm_dijet}.

Finally, the bottom right panel in Fig.~\ref{fig:ratio-dijet-monojet} shows the excluded (coloured) and allowed (grey) points in the $m_\chi$ vs.\ $m_{\zp}$ plane. The colours indicate 
which is the most constraining analysis (largest $r_{\rm obs}$) in each point of parameter space. As expected, the $\etmiss$ searches exclude the  $m_{\zp} \lesssim 1$~TeV region, while the resonance searches are more relevant at higher masses and along the $m_{\chi} \sim m_{\zp}/2$ diagonal. Important to note is the gap in coverage for $1.1 \mbox{ TeV} \lesssim m_{\zp} < 1.5 \mbox{ TeV}$ due to the fact that the ATLAS-EXOT-2019-03 (dijet) search provided limits only for $m_{\zp} > 1.5$~TeV and the other analyses are not sensitive enough to exclude this region. 

\begin{figure}[t]
    \centering
    \includegraphics[width=0.6\textwidth]{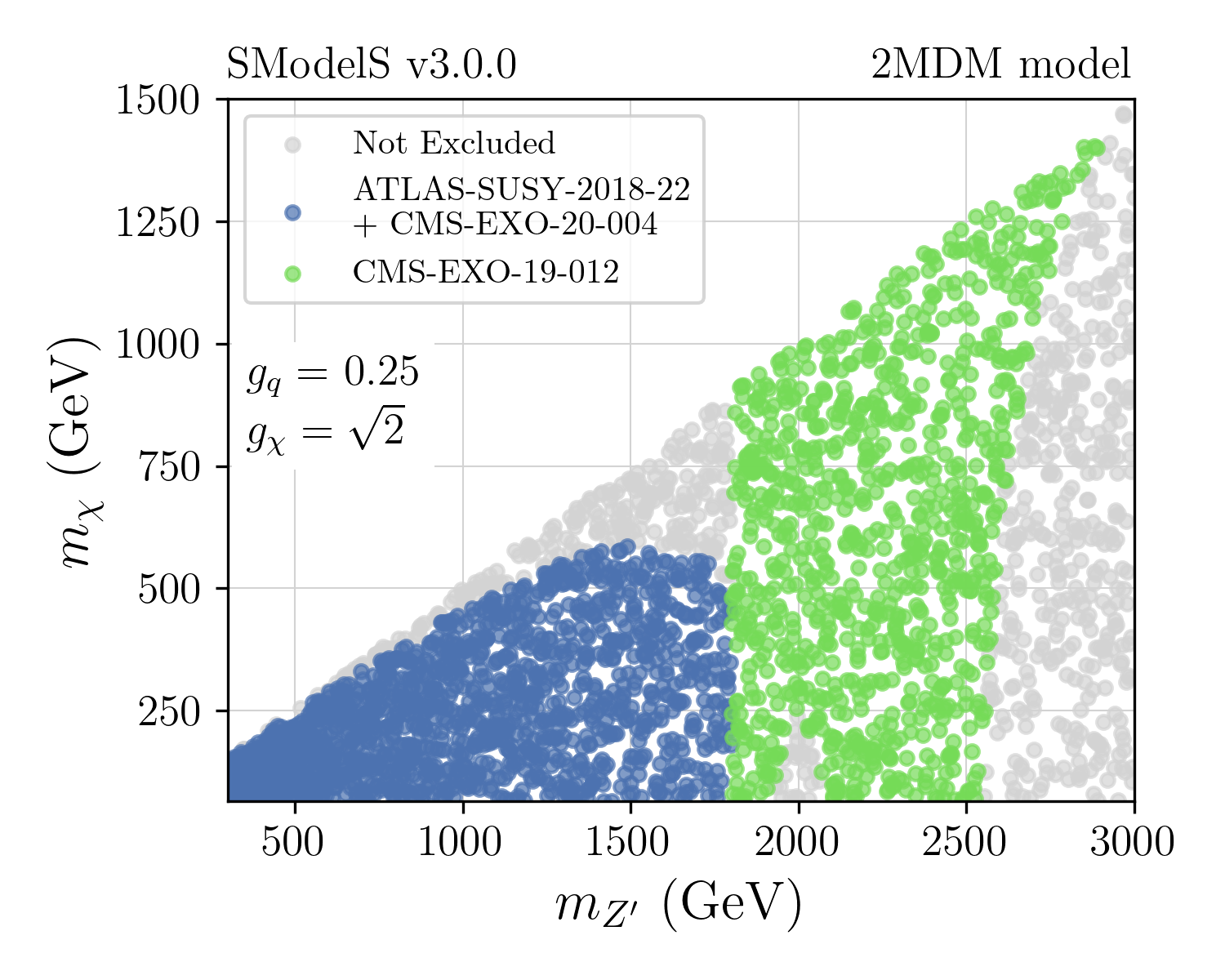}
    \caption{Regions excluded by the combination of the $\etmiss$ searches ATLAS-SUSY-2018-22 and CMS-EXO-20-004 (in blue) and by the dijet resonance search CMS-EXO-19-012 (in green) for $g_{\chi} = \sqrt{2}$ and $g_{q} = 0.25$.}
    \label{fig:dijet-monojet}
\end{figure}

Last but not least, we show in Fig.~\ref{fig:dijet-monojet} how the $\etmiss$ and resonance exclusions change
if we increase the couplings to $g_{q} = 0.25$ and $g_{\chi} = \sqrt{2}$. 
For larger couplings, the $\zp$ production cross-section increases, but so does the $\zp$ width.  
For the values chosen in Fig.~\ref{fig:dijet-monojet}, $\Gamma_{\zp}/m_{\zp}>1\%$ throughout, as we know from Fig.~\ref{fig:ratio-wm}. 
Therefore, the only applicable resonance search is CMS-EXO-19-012.
As we can see, the $\etmiss$ exclusion is larger than in Fig.~\ref{fig:ratio-dijet-monojet}, due to the larger cross-sections. The region excluded by CMS-EXO-19-012 also increases and overlaps with the combination of the $\etmiss$ searches. Overall, 
masses of $m_{\zp} < 2.5$~TeV are excluded either by the $\etmiss$ or resonance searches. This extends up to almost 3~TeV for $m_{\chi} \simeq m_{\zp}/2$. Nonetheless, a small non-excluded region 
remains for $m_{\zp} < 1.8 \mbox{ TeV}$ and $m_{\chi} \sim m_{\zp}/2$; this might be largely closed if the CMS-EXO-19-012 analysis provided results for $\zp$ masses below 1.8~TeV and/or ATLAS-EXOT-2019-03 provided width-dependent results.
Finally, we notice that a small region around $m_{\zp} \simeq 2$~TeV and $m_{\chi} \lesssim 250$~GeV escapes the dijet resonance exclusion.
This is due to a $1\sigma$ excess observed by CMS in this region, which reduces the $r_{\rm obs}$ value. 
This suppression of $r_{\rm obs}$ is also visible in the solid and dotted green curves in Fig.~\ref{fig:exclusions_2mdm_dijet} (left). 
The upward and downward fluctuations in observed data could be mitigated by combining distinct dijet searches, as it has been done for the $\etmiss$ searches.
To draw more accurate and statistically robust conclusions, however, width-dependent EM-type results and likelihoods would be needed.

\section{Conclusions}\label{sec:conclusions}

With the huge number of searches performed by the ATLAS and CMS experiments, assessing the impact of the LHC results on new physics models can be quite challenging.
Since most of the search results are currently interpreted and communicated by the experimental collaborations in terms of simplified models, \smodels aims at providing the means for translating constraints on simplified models to constraints on full UV models. 

Until \mbox{version 2}, the \smodels framework was limited to models containing a $\Ztwo$-like symmetry, such that BSM particles are pair produced and cascade decay to the lightest BSM state. 
This limitation is overcome in \smothree, presented in this paper, through the introduction of a new, graph-based topology description, allowing the tool to handle arbitrary simplified models.
In particular, searches constraining single (resonant) production of new particles, associated production of BSM and SM states, and/or decays to only SM final states (such as in RPV SUSY) can now be included in the \smodels database and used to test arbitrary BSM models.
It is worth noting here that resonance searches are particularly well-suited for reinterpretation in the simplified model approach, as, apart from the discussed width dependence, they are largely model-independent.

Despite the conceptual changes, for which the \smodels code and database were completely revised, the changes to the user API remained small, allowing an easy transition for users already familiar with previous versions of the tool.
A new version of the database was also released, including, among others, searches for resonances decaying to dijets, $b\bar{b}$, $t\bar{t}$, or DM. Also added were searches for long-lived particles decaying to multiple jets, as expected in supersymmetric RPV models. 
Altogether, the \smothree database now covers 125 ATLAS and CMS searches for new physics; 12 of these concern models without a $\Ztwo$-like symmetry.

The physics impact of the new version was illustrated through the investigation of the constraints on the Two Mediator Dark Matter model, which features a heavy $\zp$ gauge boson decaying to quarks or to DM. 
Within this model, searches for diquark resonances and $\etmiss$ searches are both relevant to cover distinct regions of parameter space. 
While a simplified version of this model is often considered by the experimental collaborations for a given set of benchmark couplings, \smothree allows us to quickly investigate how the constraints change for distinct couplings and masses.
In addition, \smodels provides the means for combining statistically independent analyses. 
This not only increases the potential physics reach, it also helps to mitigate the effect of statistical fluctuations, and to pinpoint mutually compatible excesses.

Within the 2MDM scenario, we showed that the combination of the ATLAS multijet and the CMS monojet searches provides the most sensitive constraints on the invisible $\zp$ decay,  excluding masses up to $m_{\zp} \simeq 1.8$~TeV,
for $g_q = 0.25$ and $g_{\chi} = \sqrt{2}$.
The resonance searches become relevant at higher masses, excluding the range  $1.8 \mbox{ TeV} \lesssim m_{\zp} \lesssim 2.5$~TeV for the same coupling values.
Although the use of both types of searches allows to cover a significant fraction of the parameter space, a few gaps remain within the simplified model results provided by the experimental collaborations. 
This lack of coverage is mostly due to two factors. First, most of the simplified model constraints provided by the resonance searches are
given assuming the NWA, which is only valid for small couplings. 
Second, most of the Run 2 limits start at $m_{\zp} \simeq 1.5$~TeV, hence the low mass region is only constrained by 8 TeV results, which are typically less sensitive.
The results presented here illustrate the importance of considering both narrow and broad resonances when presenting results as well as extending the results for the low mass region whenever possible.

The work presented here constitutes an important step towards a more systematic use of all the simplified model results provided by ATLAS and CMS. 
Further improvements are expected in future versions of \smodels and its database. 
An important database extension will be the addition of efficiency maps for other relevant $\etmiss$ and resonance searches, which can be obtained through recasting tools, as has been done in this work for the CMS monojet and ATLAS multijet searches.
Extended coverage by EMs will allow for a better statistical treatment, including a proper study of the compatibility of the excesses observed in some analyses with constraints from other searches. 
Also important will be the inclusion of searches for resonances decaying to pairs of leptons or gauge bosons, in order to provide a broader coverage of generic BSM models.

\section*{Acknowledgements}

We thank Rafa{\l} Mase{\l}ek for uncovering a bug in the \smodels--pyhf interface, which led to the new syntax for matching region names presented in Appendix~\ref{app:pyhfmap}.  

\paragraph{Funding information}
MMA~is supported by the French Agence Nationale de la Recherche (ANR) under grant ANR-21-CE31-0023. 
AL~is supported by FAPESP grants no.\ 2018/25225-9 and 2021/01089-1.
SN~is supported by the Austrian Science Fund (FWF) under grant number~I~5767-N\@. 
TP~is supported by the Initiatives de Recherche \`a Grenoble Alpes (IRGA) ANR-15-IDEX-02 project no.\ G7H-IRG21B26. 
CR~is supported by CAPES projects no.\ 88887.645500/2021-00 and 88887.935477/2024-00.
YV~is supported by FAPESP grants no.~2018/25225-9 and 2023/01197-4.

\appendix
\section{Additional Changes in the Database}\label{app:database}

The database~\cite{smo3:database} which comes with version~3.0.0 of \smodels features some additional changes, which were not presented in the body of this paper, because they are not directly relevant for the new graph-based topology description or the physics case study. They are thus presented in this appendix. 
This concerns additional new results described in Appendix~\ref{app:newresults}; the ability to include control regions (CRs) as well as signal uncertainties in EM-type results, described in Appendix~\ref{app:CRs}; and finally an improved syntax for mapping signal and control region names between \smodels and pyhf, described in Appendix~\ref{app:pyhfmap}. 

\subsection{Additional new results} \label{app:newresults}

\paragraph{ATLAS-SUSY-2018-09~\cite{ATLAS:2019fag}:}
This is a search in final states with same-sign leptons and jets using 139~fb$^{-1}$ of Run~2 data. It targets R-parity conserving (RPC) as well as RPV signals of gluinos and squarks. We implemented the UL maps for two simplified models: gluino-pair production followed by $\tilde g\to t\tilde t_1$, $\tilde t_1\to j b$ (RPV) and 
sbottom-pair production followed by $\tilde b\to t\tilde\chi_1^\pm$, $\tilde\chi_1^\pm \to W \tilde\chi_1^0$ (RPC).

\paragraph{ATLAS-SUSY-2018-16~\cite{ATLAS:2019lng}:}
This is a search for events with $\etmiss$ and 2 low $p_T$, opposite-sign, same-flavour leptons, using 139~fb$^{-1}$ of Run~2 data.
The requirement of jets from initial state radiation or vector-boson fusion processes allows the search to target the electroweak production of SUSY particles with compressed mass spectra.
We implemented UL- and EM-type results for $p p \to \tilde\chi_2^0 \tilde\chi_1^\pm \to Z^* W^* \tilde\chi_1^0 \tilde\chi_1^0$ for the case $m_{\tilde\chi_2^0} = m_{\tilde\chi_1^\pm}$, and for $p p \to \tilde \ell \tilde \ell \to \ell \ell \tilde\chi_1^0 \tilde\chi_1^0$. 
Since signal leakage is relevant for this analysis, we extracted EMs for both the SRs and the CRs (see Section~\ref{app:CRs} below) from the pyhf patchsets provided on HEPData~\cite{hepdata.91374.v5/r6}. The SRs and CRs are combined using the full statistical model also published on HEPData. 
In addition, we implemented the EMs for the inclusive SRs provided by ATLAS for the production of $\tilde\chi_2^0 \tilde\chi_1^\pm$, $\tilde\chi_2^0 \tilde\chi_1^0$ and $\tilde\chi_1^+ \tilde\chi_1^-$ (higgsino-like scenario), with $\tilde\chi_2^0 \to Z^* \tilde\chi_1^0$, $\tilde\chi_1^\pm \to W^* \tilde\chi_1^0$ and $m_{\tilde\chi_1^\pm} = ( m_{\tilde\chi_2^0} + m_{\tilde\chi_1^0} ) / 2$. 

\paragraph{ATLAS-SUSY-2018-33~\cite{ATLAS:2020xyo}:}
A search for long-lived particles decaying into hadrons and at least one muon using  136~fb$^{-1}$ of Run 2 data. The analysis selects events that pass a muon or $\etmiss$ trigger and contain a displaced muon track and a displaced vertex. 
The targeted scenario is the pair-production of long-lived stops that decay via a small RPV coupling into a quark and a muon. 
For the implementation in \smodels, we extracted EMs for the two SRs from the pyhf patchsets~\cite{hepdata.91760.v2/r1}, as they provide a finer binning in mass vs.\ lifetime than the acceptance and efficiency tables also available on HEPData.

\paragraph{ATLAS-SUSY-2018-42~\cite{ATLAS:2022pib}:} This is a Run~2 search for massive, charged, long-lived particles with large ionisation energy loss, using 139~fb$^{-1}$ of data. 
\smodels v2.3
already contained UL and EM results for this search for the pair production of long-lived gluinos using two inclusive SRs.
In the current version, the inclusive signal regions were split into 25 mass windows for the long lifetime ($\geq$ 1 ns) regime,
more closely reproducing the SRs adopted by ATLAS.
Moreover, in addition to the gluino scenario, EMs for these SRs were added for chargino-pair and chargino-neutralino production: $p p \to \tilde \chi_1^+ \tilde \chi_1^-$ and $p p \to \chi_1^\pm \tilde \chi_1^0$ with $\chi_1^\pm \to \pi^\pm \tilde \chi_1^0$.
These EMs were computed using the recasting code available in the \href{https://github.com/llprecasting/recastingCodes}{LLP recasting code GitHub repository}~\cite{llprepo:atlashscp}.

\paragraph{ATLAS-SUSY-2019-08~\cite{ATLAS:2020pgy}:} This is a Run~2 search for chargino-neutralino production, where the chargino decays into a $W$ boson and the LSP, while the neutralino decays into a 125~GeV SM-like Higgs boson $h$ and the LSP, based on 139~fb$^{-1}$ of data.
The analysis thus requires one lepton, $\etmiss$ and two $b$-tagged jets consistent with the decay of the SM-like Higgs boson. It has 3 inclusive and 9 exclusive SRs binned through cuts on the transverse mass and the contransverse mass $m_{\textrm T}$ and $m_{\textrm CT}$ respectively. In addition to the EMs and UL maps for $\tilde\chi_2^0 \tilde\chi_1^\pm \to h W \tilde\chi_1^0 \tilde\chi_1^0$, which were already included in previous releases, we implemented EMs for 3 new topologies based on a recast with MadAnalysis5~\cite{DVN/BUN2UX_2020}. These are for $\tilde\chi_3^0 \tilde\chi_2^\pm$ production, where:
\begin{itemize}

\item $\tilde\chi_3^0 \to h \tilde\chi_1^0$ and $\tilde\chi_2^\pm \to W^\pm \tilde\chi_2^0$,
\item  $\tilde\chi_3^0 \to h \tilde\chi_2^0$ and $\tilde\chi_2^\pm \to W^\pm \tilde\chi_1^0$,
\item $\tilde\chi_3^0 \to h \tilde\chi_2^0$ and $\tilde\chi_2^\pm \to W^\pm \tilde\chi_2^0$, 
\end{itemize}

\noindent with $\tilde\chi_2^0 \to q \bar{q} \tilde\chi_1^0$ and $( m_{\tilde\chi_2^0} - m_{\tilde\chi_1^0} )$ varying between 0 and 50~GeV.

\paragraph{CMS-SUS-21-007~\cite{CMS:2022idi}:}
This is a search in events with a single charged lepton (electron or muon), multiple hadronic jets and $\etmiss$ using 138~fb$^{-1}$ of Run~2 data. 
Using the data provided on HEPData~\cite{hepdata.135454}, UL maps were implemented for gluino pair production with $\tilde g \to \tilde\chi^0_1+t\bar t$ or $\tilde g \to \tilde\chi^0_1+q\bar q W^{(*)}$.

\subsection{Signal leaking to control regions and signal uncertainties}
\label{app:CRs}

In the \smodels-pyhf interface, by default, CRs are removed from the statistical model~\cite{Alguero:2020grj}. The possibility to prevent this has been added with \smodels v.2.2.0 through the {\tt includeCRs} flag in the {\tt globalInfo.txt} files. From v3.0.0 onward, \smodels can emulate signal leaking into the CRs if efficiency maps are provided for them. If {\tt includeCRs = True}, the statistical treatment of CRs is identical to the one of SRs, except that the former are not allowed to individually constrain the tested model. 
If {\tt includeCRs = False}, the CRs as well as the leaking signals are removed from the statistical model. 

Additionally, v3.0.0 provides the possibility to incorporate a signal uncertainty within the pyhf statistical model. To this end, we introduced a {\tt signalUncertainty} field in the {\tt globalInfo.txt} file. It allows each nominal signal of each bin of each channel (CRs as well as SRs) to be modified through an additive nuisance parameter constrained by a Gaussian; more precisely, this is done through the pyhf ``correlated shape'' modifiers. 

At the moment, these new features are used only in the ATLAS-SUSY-2018-16 analysis implementation, assuming a 22\% signal uncertainty to properly reproduce the official exclusion curve.
Generally, these choices are made at the level of the analysis implementation and are not directly accessible to the user. They can, however, be modified by the user in the \href{https://github.com/SModelS/smodels-database-release/releases}{text database}.

\subsection{Matching region names between \smodels and pyhf} \label{app:pyhfmap}

In \smodels, the naming of signal and control regions of an analysis usually follows the convention used in the experimental paper publication. However, this is not always the same as in the corresponding pyhf json file.
To unambiguously match the region names between \smodels and pyhf, the {\tt jsonFiles} entry in the {\tt globalInfo.txt} files has been extended to a dictionary format. The new format also allows us to specify which regions are SRs and which are CRs. The {\tt globalInfo.txt} of the ATLAS-SUSY-2018-04 analysis is a concrete example: 

\begin{verbatim}
id: ATLAS-SUSY-2018-04
sqrts: 13*TeV
lumi: 139.0/fb
prettyName: 2 hadronic taus
url: https://atlas.web.cern.ch/Atlas/GROUPS/PHYSICS/PAPERS/SUSY-2018-04/
....
....
jsonFiles: {
  'SRcombined.json': [
    {'pyhf': 'QCR1cut_cuts', 'type': 'CR'},
    {'pyhf': 'QCR2cut_cuts', 'type': 'CR'},
    {'smodels': 'SRlow', 'pyhf': 'SR1cut_cuts'},
    {'smodels': 'SRhigh', 'pyhf': 'SR2cut_cuts'},
    {'pyhf': 'WCRcut_cuts', 'type': 'CR'}]
  }
includeCRs: False
\end{verbatim}

As can be seen, the connection of SModelS with the pyhf model is specified as
a dictionary, with the json file name as the keys and a list of analysis region
entries as the values. The region entries match the SModelS names ({\tt smodels}) onto the pyhf region names ({\tt pyhf}) used in the json file; the region type (signal, control, or validation region) is 
specified as {\tt type} (default: {\tt 'SR'}). If the pyhf name is omitted, it is assumed to be equal to the SModelS name. If the SModelS name is omitted, we assume {\tt None} as value, indicating
that there is no corresponding EM, so this region will always have zero signal counts.
This is often the case for control regions. 
Finally, in case the region names in the pyhf model and in SModelS coincide, a simple name string can be used instead of a dictionary, as is illustrated by the ATLAS-SUSY-2018-14 example:
\begin{verbatim}
jsonFiles: {'SRee_bkgonly.json': ['SRee'], 'SRmm_bkgonly.json': ['SRmm'], 
            'Comb_bkgonly.json': ['SRee', 'SRmm', 'SRem']}
\end{verbatim}

\section{More Details on the 2MDM Model} \label{app:2mdm}

In this section, we give additional details about the 2MDM model discussed in Section~\ref{sec:model}.
Since similar versions of this model were discussed in the recent literature~\cite{Duerr:2016tmh,Argyropoulos:2021sav,Duerr:2017uap,FileviezPerez:2015mlm,Ellis:2017tkh,Butterworth:2024eyr}, we focus on the essential features relevant for the LHC results presented in Section~\ref{sec:results}. In particular, we give more details about the scalar sector of the model, the relevant Feynman rules, and the relevant decay widths for $\zp$ and $S$.

\subsection*{Scalar potential}

The 2MDM model Lagrangian is given by Eqs.~\eqref{eq:lagAll}--\eqref{eq:lagChi}, with the $\up$ charges of the SM and BSM fields listed in Table~\ref{tab:u1charges}. $\mathcal{L}_{\phi}$, Eq.~\eqref{eq:lagS},  contains the scalar potential for the new scalar $\phi$,
\begin{equation}
V_{H,\phi} = \mu_{2}^2 |\phi|^2 + \lambda_2 |\phi|^4 + \lambda_3 |\phi|^2 |H|^2 \,,
\end{equation}
where $H$ represents the SM Higgs doublet and we define the SM Higgs potential as
\begin{equation}
V_{H} = \mu_1^2 |H|^2 + \lambda_1 |H|^4 \,.
\end{equation}
Assuming that both $H$ and $\phi$ develop vevs, $\langle \phi \rangle = v_2/\sqrt{2}$ and $\langle H \rangle = v/\sqrt{2}$, imposes the following minimization conditions on the scalar potential parameters:
\begin{equation}
    \mu_1^2 =  -\left(\lambda_{1}v^2 + \lambda_{3} v_2^2 \right), 
    \quad
    \mu_2^2 = -\left(\lambda_{2} v_2^2 + \lambda_{3}\frac{v^2}{2}\right) 
    \label{eq:mu}
\end{equation}
and
\begin{equation}
    4  \lambda_1 \lambda_2- \lambda_3^2 > 0\mbox{, } \lambda_1 \lambda_2 >0 \,.
\end{equation}

After electroweak symmetry breaking (EWSB), the neutral components of the SM Higgs doublet ($H^0$) and the singlet ($\phi^0$) mix, resulting in the mass eigenstates 
$h = H^0 \cos \alpha - \phi^0 \sin \alpha $ and 
$S = \phi^0 \cos \alpha + H^0 \sin \alpha $, cf.~Eq.~\eqref{eq:alpha}.  
The masses are, in terms of $\lambda$'s, $v$, $v_2$ and $\alpha$,    
\begin{equation}
m_{S,h}^2 = \lambda_{1} v^2+\lambda_{2} v_2^2 \mp \left(\lambda_{1} v^2-\lambda_{2} v_2^2\right) \sqrt{1 + \tan^2(2 \alpha)}\,,
   \label{mhs}
\end{equation}
with the mixing angle given by
\begin{equation}
    \tan(2 \alpha) \equiv \frac{\lambda_{3} v v_2}{\lambda_1 v^2-\lambda_2 v_2^2}\,.
    \label{tan}
\end{equation}
Using the above equations, we can express the quartic couplings ($\lambda_i$) in terms of the masses, mixing angle and vevs:  
\begin{align}
 \lambda_1 &=  \frac{1}{2 v^2}\left(\cos^2\alpha\, m_{h}^2+m_{S}^2 \sin^2\alpha\right),\\
 \lambda_2 &= \frac{2 g_{\chi}^{2}}{m_{\zp}^2} \left(\cos^2\alpha \, m_{S}^2+ m_{h}^2 \sin^2 \alpha\right),\\
  \lambda_{3} &= \frac{2 g_{\chi}}{m_{\zp} v} \left(m_{S}^2 -m_h^2 \right) \sin\alpha \cos\alpha \; . \label{eq:lam3} 
\end{align}
Therefore, the parameters of the scalar potential ($\mu_{1,2}$ and $\lambda_{1,2,3}$) can be replaced by $m_{h},m_S,v,v_2$ and $\sin\alpha$. Finally, $v_2$ can be replaced by $m_{\zp}$, since $v_2 = m_{\zp}/(2 g_{\chi})$, with $g_{\chi} = g_{\zp} q_{\chi}$.

\subsection*{Feynman rules}

The relevant Feynman rules for the BSM particles are given in \cref{tab:feynmanRules} below. Here, $f$ represents any of the SM fermions and $q$ any SM quark.

\begin{table}[h!]   \centering
    \rowcolors{2}{gray!10}{white}
    \vspace{0.2cm}
        \caption{Feynman rules for the relevant interactions of the $\zp$, $S$ and $\chi$. \label{tab:feynmanRules}}
    \begin{tabular}{p{2cm}|p{8.5cm}}
      \toprule
      \textbf{Interaction} & \textbf{Vertex term}\\ \toprule 
      $\zp_\mu \,q\,\bar{q}$ & $i g_q \gamma^\mu$\\
      $\zp_\mu\,\chi\,\chi$ & $- i g_{\chi} \gamma^\mu \gamma^5$\\
      $S\,f\,\bar{f}$  & $-i \frac{m_f}{v} \sin\alpha$ \\
      $S\,\chi\,\chi$  & $ - 2i g_{\chi} \frac{m_{\chi}}{m_{\zp}} \cos\alpha$\\
      $S\,W_\mu^-\,W_\nu^+$  & $2 i g^{\mu\nu} \frac{m_{W}^2}{v} \sin\alpha$\\
      $S\,Z_\mu\,Z_\nu$  & $2 i g^{\mu\nu} \frac{m_{Z}^2}{v} \sin\alpha$\\
      $S\,h\,h$  & $- i \frac{m_{S}^{2}}{2 m_{\zp} v}  \left( 1 + 2 \frac{m_{h}^{2}}{m_{S}^{2}}\right)  \left( m_{\zp} \cos\alpha + 2 g_{\chi} v \sin\alpha \right)  \sin (2\alpha)$\\
      $S\,\zp_\mu\,\zp_\nu$  &  $4 i g^{\mu \nu} g_{\chi} m_{\zp}  \cos \alpha$\\\bottomrule        
    \end{tabular}
\end{table}

\subsection*{Decay widths}

Using the Feynman rules from \cref{tab:feynmanRules}, we can easily compute the 2-body decay widths for $\zp$ and $S$. 
The partial widths of the $\zp$ are directly proportional to the $\up$ charges of the SM and DM particles and read:
\begin{eqnarray}
    \Gamma(\zp \to q  \bar{q}) & = &  \frac{{g_{q}^{2} m_{\zp}}}{4 \pi} \sqrt{1- \frac{4 m_{q}^{2}}{m_{\zp}^{2}}} \left(1 + \frac{2 m_{q}^{2}}{m_{\zp}^{2}}\right) , \\
    \Gamma(\zp \to \chi \chi) & = &  \frac{g_{\chi}^{2}m_{\zp}}{24 \pi} \left(1 - \frac{4 m_{\chi}^{2}}{m_{\zp}^{2}}\right)^{\sfrac{3}{2}} . 
\end{eqnarray}
Regarding the decays of $S$, since the scalar $\phi$ field is a singlet under the SM gauge groups, $S$ decays to SM particles proceed only through the mixing with the Higgs, which is  suppressed by $\sin\alpha$. 
The couplings to $\zp$ and $\chi$, on the other hand, are proportional to $\cos\alpha$. Since here we assume $m_{\chi} < m_S < m_{\zp}$, we have the following decay widths:
\begin{eqnarray}
\Gamma(S \to \ell \bar{\ell} ) & = &  \frac{m_{S}}{16 \pi} \frac{m_{l}^2}{v^2} \left(1 - \frac{4 m_{\ell}^{2}}{m_{S}^{2}}\right)^{\sfrac{3}{2}} \sin^{2}\alpha \,,  \\
\Gamma(S \to  q \bar{q} ) & = &   3\frac{ m_{S}}{16\pi}\frac{m_q^2}{v^2} \left(1 - \frac{4 m_q^{2}}{m_{S}^{2}}\right)^{\sfrac{3}{2}} \sin^{2}\alpha \,, \\
\Gamma(S \to \chi  \chi  ) &= &  g_{\chi}^2\frac{m_{S}}{4 \pi} \frac{m_\chi^2}{m_{\zp}^2}\left(1 - \frac{4 m_{\chi}^{2}}{m_{S}^{2}}\right)^{\sfrac{3}{2}} \cos^{2}\alpha \,, \\
\Gamma(S \to W W ) &= &  \frac{m_{S}^{3}}{16 \pi v^{2}} \sqrt{1 - \frac{4 m_{W}^{2}}{m_{S}^{2}}} \left(1 - \frac{4 m_{W}^{2}}{m_{S}^{2}} + \frac{12 m_{W}^{4}}{m_{S}^{4}}\right) \sin^{2}\alpha \,, \\
\Gamma(S \to Z Z ) &= &  \frac{m_{S}^{3}  }{32\pi v^{2}} \sqrt{1 - \frac{4 m_{Z}^{2}}{m_{S}^{2}}}  \left( \frac{4 m_{Z}^{2}}{m_{S}^{2}} - \frac{12 m_{Z}^{4}}{m_{S}^{4}} - 1\right) \sin^{2}\alpha \,, \\
\Gamma(S \to h h ) & = & \frac{m_{S}^{3}}{128 \pi  v^{2} m_{\zp}^{2}}  \left( 1 + 2 \frac{m_{h}^2}{m_{S}^2} \right)^{2} \sqrt{1-\frac{4 m_{h}^2}{m_{S}^2}}  \nonumber \\
& & \hspace*{2cm}\times \left( m_{\zp} \cos\alpha + 2 g_{\chi} v \sin\alpha \right)^2  \sin^2 (2\alpha) \,,
\end{eqnarray}
where $\ell$ represents any SM charged lepton and $q$ any SM quark.

\clearpage
\bibliographystyle{JHEP}
\bibliography{references}

\end{document}